\documentclass[12pt,fleqn]{article}
\usepackage[DIV12]{typearea}
\usepackage{amsmath,amsfonts,amssymb}
\usepackage{graphicx}
\usepackage{cite}
\usepackage{mathrsfs}
\usepackage{bbm}
\usepackage{pifont}
\usepackage{cancel}
\usepackage{units}
\usepackage[small,loose]{subfigure}  
\usepackage[dvipsnames]{xcolor}
\usepackage{mathtools}
\usepackage[bookmarks=false]{hyperref}
\usepackage{slashed} 
\usepackage{colortbl}
\usepackage[normalem]{ulem}
\usepackage{pdfcomment}
\usepackage{tikz}
\usepackage{float}
\restylefloat{table}
\restylefloat{figure}

\usetikzlibrary{positioning,shapes,shadows,arrows}

\newcommand{\Secref}[1]{Section~\ref{#1}}
\newcommand{\Eqref}[1]{Equation~\eqref{#1}}

\newcommand{\Tabref}[1]{Table~\ref{#1}}

\newcommand{\eVdist}{\kern-0.06em}

\newcommand{\rep}[1]{\ensuremath\boldsymbol{#1}}
\newcommand{\crep}[1]{\ensuremath\overline{\boldsymbol{#1}}}

\DeclareMathOperator{\diag}{diag}

\DeclareMathOperator{\rank}{rank}

\newcommand{\I}{\mathrm{i}}

\newcommand{\SO}[1]{\ensuremath{\mathrm{SO}(#1)}}
\newcommand{\SU}[1]{\ensuremath{\mathrm{SU}(#1)}}
\newcommand{\U}[1]{\ensuremath{\mathrm{U}(#1)}}
\newcommand{\Z}[1]{\ensuremath{\mathbbm{Z}_{#1}}} 

\newcommand{\Hu}{\ensuremath{H_u}}
\newcommand{\Hd}{\ensuremath{H_d}}
\newcommand{\qHu}{\ensuremath{q_{\Hu}}}
\newcommand{\qHd}{\ensuremath{q_{\Hd}}}

\newcommand{\DHL}[0]{DHL~\cite{Dreiner:2012ae}}

\allowdisplaybreaks[1]

\numberwithin{equation}{section}
\numberwithin{table}{section}

\def\mytitle{$R$ parity violation from discrete $R$ symmetries}

\title{\mytitle}

\begin{document}

\begin{titlepage}

\begin{flushright}
UCI--TR--2014--08\\
TUM--HEP 959/14\\
NSF--KITP--14--130\\
FLAVOUR--EU 85
\end{flushright}

\vspace*{1.0cm}

\begin{center}
{\Large\bf\boldmath\mytitle\unboldmath}

\vspace{1cm}

\textbf{
Mu--Chun Chen\footnote[1]{Email: \texttt{muchunc@uci.edu}}{}$^a$,
Michael Ratz\footnote[2]{Email: \texttt{michael.ratz@tum.de}}{}$^b$,
Volodymyr Takhistov\footnote[3]{Email: \texttt{vtakhist@uci.edu}}{}$^a$
}
\\[5mm]
\textit{\small
{}$^a$
Department of Physics and Astronomy, University of California,\\
~~Irvine, California 92697--4575, USA
}
\\[5mm]
\textit{\small
{}$^b$ Physik--Department T30, Technische Universit\"at M\"unchen, \\
~~James--Franck--Stra\ss e~1, 85748 Garching, Germany
}

\end{center}

\vspace{1cm}

\begin{abstract}
 We consider supersymmetric extensions of the standard model in which the usual
$R$ or matter parity gets replaced by another $R$ or non--$R$ discrete symmetry
that explains the observed longevity of the nucleon  and solves the $\mu$
problem of MSSM. In order to identify suitable symmetries, we develop a novel
method of deriving the maximal $\Z{N}^{(R)}$ symmetry that satisfies a given set
of constraints. We identify $R$ parity violating (RPV) and conserving
 models that are consistent with precision gauge
unification and also comment on their compatibility with a unified gauge
symmetry such as the Pati--Salam group. Finally, we provide a
counter--example to the statement found in the recent literature that the lepton
number violating RPV scenarios must have $\mu$ term and the bilinear $\kappa \,
L \, \Hu$ operator of comparable magnitude.
\end{abstract}

\end{titlepage}
\pagenumbering{arabic}


\section{Introduction}

Low--energy supersymmetry is still one of the most attractive schemes for
physics beyond the standard model (SM). One of the striking features of
supersymmetry is that it leads to precision gauge unification in the minimal
supersymmetric extension of the SM, the MSSM. Supersymmetry allows for the
stabilization of the electroweak scale against the grand unification scale,
$M_\mathrm{GUT}$, where the gauge couplings unify.  The non--observation of the
superpartners so far at the  Large Hadron Collider (LHC)
\cite{Aad:2011hh,Chatrchyan:2013sza} has placed significant constraints  on the
minimal supersymmetric models that have been generally considered.  $R$ parity
violation (RPV) turns out to be an interesting alternative
\cite{Aulakh:1982bx, Hall:1983id,Ellis:1984gi,Ross:1984yg,Dreiner:1997uz,Barbier:2004ez} to
consider beyond the minimal models,  since it may explain why the superpartners
have not been seen at the LHC (see e.g.\ \cite{Graham:2014vya} for a recent
analysis). On the other hand, in the presence of $R$ parity violation, one
should explain non--observation of nucleon decay thus far
\cite{Kobayashi:2005pe}. To achieve these two goals simultaneously requires,
naturally,  additional symmetries, with discrete symmetries being a
plausible option. Alternatives include invoking minimal flavor
violation \cite{Smith:2008ju,Csaki:2011ge}.

$R$ symmetries play a special role in this context, since the order parameter
for $R$ symmetry breaking is the gravitino mass
$m_{\nicefrac{3}{2}}$.\footnote{Recall that $R$ parity is actually not an
$R$ symmetry. Rather, it is equivalent to matter parity (see e.g.\
\cite{Chen:2012tia} for a discussion).} Thus, without having to go into details
of supersymmetry breaking, it is possible to estimate the amount by which
discrete $R$ symmetries are broken. In turn, this enables one to make statements
about the coefficients of the effective operators that arise through $R$
symmetry breaking. Such effective operators will be suppressed by powers of the
ratio of gravitino mass and the fundamental scale, 
$m_{\nicefrac{3}{2}}/\Lambda$. In general, (discrete) $R$ symmetries are broken
by the vacuum expectation value (VEV) of some ``hidden sector'' superpotential,
which carries $R$ charge $2q_\theta$, where $q_\theta$ denotes the $R$ charge of
the superspace coordinate $\theta$, and possibly by further operators. This
allows for the possibility of residual non--$R$ $\Z{M}$ symmetries, in
particular for $q_\theta>1$ \cite{Chen:2012jg}. In light of arguments from
quantum gravity \cite{Krauss:1988zc}, we will focus on gauged discrete
symmetries.

In a recent analysis, Dreiner, Hanussek and Luhn (DHL)~\cite{Dreiner:2012ae}
analyzed discrete $R$ symmetries of the type described above.  In their work,
the $R$ charge of the superspace coordinate $\theta$ was restricted to 1.
Further, \DHL\ allowed for the Green--Schwarz (GS) mechanism~\cite{Green:1984sg}
to cancel the anomalies,  and required that the couplings of the GS axion $a$ to
the various field strengths of the (MS)SM be universal. On the other hand, as
pointed out in \cite{Ludeling:2012cu}, universality of the anomaly coefficients
is, strictly speaking, not a consistency condition, even though one may impose
it in order not to spoil precision gauge unification \cite{Chen:2012tia}.
Precision gauge unification may also be preserved, for example, if the scalar
partner of the axion $a$ has an expectation value that is much smaller than the
axion decay constant $f_a$, or by an accidental cancellation of unrelated
effects \cite{Dine:2012mf}.

The purpose of this work is to complete and to extend the analysis of \DHL\ by
\begin{itemize}
 \item allowing the superspace coordinate $\theta$ to have an $R$ charge that
 differs from 1;
 \item allowing for a GS cancellation of discrete anomalies with non--universal
 couplings of $a$;
\item identifying redundant symmetries in \DHL;
\item presenting a novel method allowing one to systematically identify the
maximal symmetry compatible with given selection criteria.
\end{itemize}
Moreover, we will also comment on $R$ parity conserving scenarios. 

In our analysis, we consider both $R$ and non--$R$ Abelian discrete symmetries,
and impose that
\begin{enumerate}
 \item the nucleon is sufficiently long--lived, i.e.\ that the dangerous
 operators are either forbidden by a residual $\Z{M}$ symmetry or sufficiently
 suppressed by appropriate powers of $m_{\nicefrac{3}{2}}/\Lambda$. Here,
 $\Lambda$ is the cutoff scale which we take to be the Planck scale
 $M_\mathrm{P}$ unless stated otherwise. We will also discuss settings with
 a lower cutoff scale.
 \item the discrete symmetry forbids the $\mu$ term at the perturbative level.
\end{enumerate}
Further, we demonstrate additional features that were absent from \DHL\, including
\begin{itemize}
 \item the compatibility of charges with (partial) unification, specifically
whether the matter charges commute with the Pati--Salam group
$G_\mathrm{PS}=\SU4\times\SU2_\mathrm{L}\times\SU2_\mathrm{R}$;\footnote{We do
not consider compatibility of matter charges with \SU5 or \SO{10} in the case of
RPV. This is because, if $\overline{U}\,\overline{D}\,\overline{D}$ is allowed,
so is automatically $L\,L\,\overline{E}$, and vice versa. See
\cite{Acharya:2014vha} for a discussion of $R$ parity violation in settings with
GUT relations.}
 \item  a natural suppression of the neutrino masses either through the Weinberg
 operator or from supersymmetry breaking, thus yielding light
 Dirac neutrinos.
\end{itemize}

This paper is organized as follows. In \Secref{sec:classify}, we present a novel
method for classifying discrete symmetries. We comment on anomaly cancellation, 
provide a recipe for identifying and eliminating equivalent symmetries, and  
comment on the limitations of our analysis. In
\Secref{sec:models}, we illustrate our methods by presenting models obtained for
anomaly--universal as well as non--universal scenarios while considering both
$R$ parity violation and conservation.  \Secref{sec:summary} contains our
conclusions.


\section{Classification}
\label{sec:classify}

\subsection{Goals of our classification}
\label{sec:strategy}

In the MSSM, the renormalizable superpotential terms consistent with the SM
gauge symmetry are
\begin{align}		
\mathscr{W}_\mathrm{ren} ~ =~ &
 \mu\, \Hu\Hd
 +Y^u_{fg}\,Q_f\,\overline{U}_g\,\Hu
 + Y^d_{fg}\,Q_f\,\overline{D}_g\,\Hd
 + Y^e_{fg}\,L_f\,\overline{E}_g\,\Hd
 \notag\\
 &{}
 +\kappa^f\, L_f\,\Hu
 +\lambda^{fgh}\, L_f\,L_g\,\overline{E}_h
 +\lambda^{\prime\,fgh}\,L_f\,Q_g\,\overline{D}_h
 +\lambda^{\prime\prime\,fgh}\,\overline{U}_f\,\overline{D}_g\,\overline{D}_h
 \;,
\end{align}
where the first line denotes the usual couplings of the MSSM, while the
second line contains the so--called $R$ parity violating terms. In what follows,
we will suppress the flavor indices $f$, $g$ and $h$. We will further assume
that there is no flavor dependence of the discrete charges, i.e.\ $q_Q^f=q_Q$
for all $f$ and so on.

At the non--renormalizable level, additional $B$ 
and $L$ violating operators need to be considered (cf.\ e.g.\
\cite{Ibanez:1991pr, Allanach:2003eb, Dreiner:2005rd, Dreiner:2012ae})
\begin{align}
 \label{eq:HigherDimOperators}
 && \mathcal{O}_1 &~=~\left[Q\,Q\,Q\,L\right]_F \;, 			          &
 \mathcal{O}_2 &~=~\left[\overline{U}\, \overline{U}\, \overline{D}\, \overline{E}\right]_F\;,				&&  \notag\\
 && \mathcal{O}_3 &~=~\left[Q\,Q\,Q\,\Hd\right]_F\;,   	          &
 \mathcal{O}_4 &~=~\left[Q\, \overline{U}\, \overline{E}\,\Hd\right]_F	\;,			    &&   \notag\\
 && \mathcal{O}_5 &~=~\left[L\,\Hu\, L\,\Hu\right]_F \;, 
 &  \mathcal{O}_6 &~=~\left[L\, \Hu\, \Hd\, \Hu\right]_F	\;,                  &&  \notag 		\\
&& \mathcal{O}_7 &~=~\left[\overline{U}\, \overline{D}^{\dagger}\, \overline{E}\right]_D\;,
&  \mathcal{O}_8 &~=~\left[\Hu^{\dagger}\,  \Hd\, \overline{E}\right]_D	\;,              &&   \notag\\
 && \mathcal{O}_9 &~=~\left[Q\, \overline{U}\, L^{\dagger}\right]_D  \;,		 &   
 \mathcal{O}_{10} &~=~\left[Q\,Q\,\overline{D}^{\dagger}\right]_D	\;,	            	&&
\end{align}
as well as operators of even higher dimensions.

We will discuss settings with renormalizable baryon number violation
($\cancel{B}$), renormalizable lepton number violation ($\cancel{L}$) as
well as ``non--perturbative'' $B$ and $L$ violation, which appears
only after the ``hidden sector'' superpotential acquires its VEV. We will
further comment on settings with $R$ parity conservation. To constrain overly
rapid proton decay, renormalizable $\cancel{B}$ operators must be forbidden in
the case of the RPV setting with renormalizable $\cancel{L}$, and vice versa for
the RPV setting with renormalizable $\cancel{B}$. Since not all of the above
higher--dimensional operators shown in \Eqref{eq:HigherDimOperators} are
independent (see \cite{Dreiner:2005rd, Dreiner:2012ae,Ibanez:1991pr}), only a
subset of such terms need to be considered to account for all the
phenomenological constraints. In RPV setups with either renormalizable
$\cancel{B}$ or renormalizable $\cancel{L}$, one needs only to examine the
existence condition for the Weinberg operator
$\mathcal{O}_{5}$~\cite{PhysRevLett.43.1566} for neutrino mass generation. 

We consider different classes of models based on
 Abelian discrete $R$  or non--$R$  symmetries, $\Z{N}^{(R)}$, 
with properties specified below.
We distinguish between $\Z{N}^{(R)}$ symmetries 
that are anomaly--free in the traditional sense
and symmetries in which the anomalies are cancelled by a non--trivial (discrete)
Green--Schwarz (GS) mechanism \cite{Green:1984sg}. In the second case, we
discriminate between universal and non--universal couplings of the GS axion to
the various field strengths of the standard model gauge group
$G_\mathrm{SM}=\SU3_\mathrm{C}\times\SU2_\mathrm{L}\times\U1_Y$.

To sum up, we search for, both in the anomaly--universal case and in the anomaly
non--universal case, classes of models that have the following respective
properties:
\begin{enumerate}
 \item $R$ parity conservation;
 \item renormalizable RPV with $\cancel{L}$ and the existence of
  $\mathcal{O}_{5}$ at the perturbative level;
 \item renormalizable RPV with $\cancel{B}$ and the existence of
  $\mathcal{O}_{5}$ at the perturbative level;
 \item ``non--perturbative'' $\cancel{L}$ and $\cancel{B}$.
\end{enumerate}

\subsection{Equivalent discrete symmetries}
\label{sec:EquivalentDiscreteSymmetries}

In order to avoid an unnecessary double--counting of symmetry patterns,
we provide a recipe that allows to identify and eliminate equivalent
symmetries.\footnote{One possible, ``brute force'' way of doing this consists of
comparing the Hilbert bases for the K\"ahler potential and superpotential of the two
candidate symmetries. However, this turns out to be often impractical.}
This can be achieved by avoiding the following  redundancies in the definition
of the discrete $\Z{N}^{(R)}$ charges:
\begin{description}
 \item[Common divisors:]
If the order $N$ and all charges have a common divisor $M$, then the
 $\Z{N}^{(R)}$ is equivalent to a $\Z{N/M}^{(R)}$ with all charges
 divided by $M$.
 \item[Non--trivial centers:]
In the presence of an $\SU{M}$ gauge factor,
  acting with the center of \SU{M}, $Z_{\SU{M}}~\simeq~\Z{M}$, is always a
  symmetry. Thus, in the context of the standard model gauge symmetry,
  we can apply
  \begin{itemize}
  \item the non--trivial elements of the center $Z_{\SU3_\mathrm{C}}
  ~\simeq~\Z{3}$, which acts as $\diag(\omega,\omega,\omega)$ with
  $\omega=\mathrm{e}^{2\pi\,\I/3}$
  or $\mathrm{e}^{4\pi\,\I/3}$ 
  on the triplets.
  Hence, if 3 divides the order $N$, i.e.\ if $N=3\cdot N^{\prime}$,
with $N^{\prime} \in \mathbbm{N}$, this allows us
 to shift the charges according to
$(q_Q, \; q_{\overline{U}}, \; q_{\overline{D}}) \to
 (q_Q+N^{\prime}\cdot\nu, \; q_{\overline{U}}-N^{\prime}\cdot\nu, \; q_{\overline{D}}-N^{\prime}\cdot\nu)$ with
 $\nu\in\mathbbm{Z}$.
 \item  the non--trivial center of \SU2, $Z_{\SU2}
 ~\simeq~\Z{2}$. That is, equivalent charges are obtained
by multiplying all doublets by $-1$, or adding $N/2$ to the doublet
 charges, if the order $N$ is even. 
 \end{itemize}
We note that in the settings compatible with \SU5 unification there is a
 ``fake'' \Z5 which is nothing but the non--trivial center of
 \SU5~\cite{Csaki:1997aw,Petersen:2009ip}.
 \item[Hypercharge shift:] One may add integral multiples of hypercharge
 (normalized to integer charges), arriving at an equivalent charge assignment.
The freedom of performing the hypercharge shift and modding out
 the non--trivial centers of $\SU3_\mathrm{C}$ and $\SU2_\mathrm{L}$ are not independent.
 \item[Coprime factors:] Multiplying all charges by a factor $f$ that is
coprime to the order $N$, i.e.\ $\gcd(f,N)=1$, leads to the same symmetry.
Based on this, one can show that all non--$R$ symmetries of a given order in tables~2 and 3 in
\DHL\ are equivalent. That is, rather than having 2, 3, 2, 3 or 2 non--$R$ symmetries
for orders  5, 7, 8, 9 or 10, respectively, there is only one non--$R$ symmetry
for each order in table 2. Similar statements apply to the non--$R$
symmetries of table~3. We list the truly inequivalent symmetries in 
\Tabref{tab:Bviolation} and \Tabref{tab:Lviolation}.
\end{description}

The last statement also implies that for $R$ symmetries of prime order, one only
needs to consider the cases $q_\theta=0$ (which corresponds to a non--$R$
symmetry) or $q_\theta=1$. This follows from the fact that the linear congruence 
\begin{equation}
 q_\theta\cdot f~=~1\mod N
\end{equation}
has, according to the discussion in Appendix~\ref{app:Congruence}, a
non--trivial solution with a non--trivial $f$ that is coprime to $N$. More
generally, for a given order $N$, one has to scan only the values of $q_\theta$ that
divide $N$, since any other $q_\theta\ne0$ can be rescaled to 1. Further, the
case $q_\theta=N/2$ for even $N$  should not be considered. This is because the
transformation under which  $\theta\to-\theta$ and all fermions being mapped to
minus themselves is a symmetry of any supersymmetric theory.
Consequently, imposing such a transformation as a symmetry does not
forbid any couplings.

\subsection{\boldmath Systematic search for $\Z{N}^{(R)}$ 
symmetries\unboldmath}
\label{sec:SystematicSearch4ZN}

Very often in model building one encounters the situation in which one wishes
to forbid certain operators, such as some of the $\mathcal{O}_i$ in
\eqref{eq:HigherDimOperators}, by an appropriate discrete symmetry. In most
approaches, the desired symmetries and charges were found by a scan. In what
follows, we will discuss a method to systematically construct $\Z{N}$ symmetries
which allow for certain desired operators and forbid other undesired operators.

Suppose we have $n_c$ constraints, which 
correspond to $n_c$ conditions of the type
\begin{equation}\label{eq:LinearConstraints}
 \sum_{j=1}^{n_q} a_{ij}\,q_j~=~0\mod N\qquad\forall~1\le i\le n_c
\end{equation}
on the $n_q$ charges $q_j$. 
Here we concentrate first on the case with constraints of the equality
form.  Constraints in the form of inequalities will be discussed later. There
are two questions to be addressed: whether the conditions can be solved in a
non--trivial way, and, if continuous symmetries are not available, what is the
maximal meaningful $\Z{N}$ symmetry that one can impose that fulfills the
constraints. As we shall see, using the Smith normal form, which has been
shown~\cite{Petersen:2009ip,Ivanov:2013bka} to be an important tool in other
applications of discrete symmetries in physics, one can find the maximal order
$N$ of the corresponding meaningful symmetry.

Let us start by clarifying what we mean by a ``meaningful'' symmetry. Consider
a field $\phi$ transforming under a $\Z{N}$ symmetry with $\Z{N}$ charge $q$,
i.e.\
\begin{equation} \label{eq:fieldtransform}
 \phi~\xmapsto{\Z{N}}~\mathrm{e}^{2\pi\,\I\,q/N}\,\phi\;.
\end{equation}
The task is now to find a ``meaningful'' order $N$ and charge $q$ in the
1--dimensional version of \eqref{eq:LinearConstraints}, i.e.\ in the
constraint equation
\begin{equation}\label{eq:ModCondition}
 a\cdot q~=~0\mod N\;.
\end{equation}
We may rephrase this as the problem of finding the maximal meaningful symmetry
$\Z{N}$ and charge $q$, such that the operator $\phi^a$ is $\Z{N}$ invariant.
The order $N$ is a priori unknown. However, it is evident that $N=a$ 
with $q=1$. If we were to choose $N<a$, then for any integer $q$ 
 which satisfies \eqref{eq:ModCondition}, there would be
a power $a'$ such that $\phi^{a'}$ is \Z{N} invariant. That is, in addition to
the operator $\phi^{a}$, there will be additional operator(s)  $\phi^{a'}$ with
$a' < a$ allowed by the \Z{N} symmetry. Hence, we should have started from
\eqref{eq:ModCondition} with $a$ replaced by $a'$. 
On the other hand, choosing $N>a$  does not add anything useful, but will require
solutions to have $q \neq 1$. 
 That is, we would have a redundancy and not a ``meaningful''  symmetry.

Let us now look at a situation where there is another field with charge
$\widetilde{q}$, fulfilling
\begin{equation}\label{eq:ModConditionPrime}
 \widetilde{a}\cdot \widetilde{q}~=~0\mod N\;.
\end{equation}
Using analogous arguments as above, it is straightforward to convince oneself
that the maximal meaningful order is then
$N=a\cdot\widetilde{a}/\gcd(a,\widetilde{a})$. 

These statements are almost trivial and can be straightforwardly generalized to
multiple conditions of the type of \eqref{eq:ModCondition}.
A slightly more interesting situation arises when the constraints involve more
than one charge at the same time, as in \eqref{eq:LinearConstraints}. The
strategy of the subsequent discussion will be to transform these equations into
constraints on linear combinations of charges which are all of the form
\eqref{eq:ModCondition}.

Let us now discuss in detail how this works. We start out by considering
equalities of the type
\begin{equation}
 q_Q+q_{\overline{U}}+ q_{\Hu}-2q_\theta~=~0\mod N\;,
\end{equation}
a condition for the $u$--type Yukawa coupling to be allowed. Apart from the charges, the
order $N$ of the discrete symmetry is, as before, unknown.

We can rewrite the conditions of this type as
\begin{equation}\label{eq:MatrixEquation}
 A\cdot q~=~0\mod N\;,
\end{equation}
where $q$ denote the vector of the $n_q$ charges and $A$ is an integer
matrix. If $A$ does not have full rank, then there is at least one \U1 which one
can impose in order to satisfy the conditions. The \U1 charges are given by the
entries of (one of) the vector(s) in the kernel of $A$. In this case, one can
impose an arbitrary $\Z{N}$ which is a subgroup of such a \U1.  We therefore
specialize here on the case where $n_c\ge n_q$ and $A$ has full rank, such that
there is no \U1 which one may impose. Note that $A$ is not necessarily a square
matrix, i.e.\ we also allow for more constraints than variables, $n_c>n_q$. $A$
can be brought to the so--called Smith normal form,\footnote{A
\texttt{mathematica} package to compute the Smith normal form for integer
matrices can be found at
\href{http://library.wolfram.com/infocenter/MathSource/6621/}{\texttt{http://library.wolfram.com/infocenter/MathSource/6621/}}.}
\begin{equation}\label{eq:SmithForm}
 U\cdot A\cdot V~=~D\;,
\end{equation}
where $U$ and $V$ are unimodular $n_c\times n_c$ and $n_q\times n_q$ matrices,
respectively.\footnote{Recall that unimodular matrices are integer matrices with
determinant $\pm1$. The inverses of such matrices are also integer.} $D$ is also
an integer matrix and diagonal (but not necessarily square),
\begin{equation}\label{eq:MatrixD}
 D~=~\left(\begin{array}{c}
 \begin{array}{ccc}d_1 &  & \\
  & \ddots & \\
  &  & d_{n_q}
 \end{array}\\
 \hline
 0_{(n_c-n_q) \times n_q}
 \end{array}\right)
 \;,
\end{equation}
and the diagonal elements satisfy $d_i|d_{i+1}$, i.e.\ the $i^\mathrm{th}$
element divides the $(i+1)^\mathrm{th}$ element. It is also possible that
$d_{i\ge n_0}=0$ for some $n_0\le n_q$. If $D$ has maximal rank, and if it was
not for the modulo $N$, then the matrix equation \eqref{eq:MatrixEquation} has
only the trivial solution. However, if we only ask the conditions to be
fulfilled modulo $N$, then the last non--trivial element $d_{n_q}$ defines the
maximal order of a meaningful $\Z{N}$ symmetry, i.e.\ $N=d_{n_q}$, under which
the conditions encoded by \eqref{eq:MatrixEquation} can hold for non--trivial
$q$.  This can be seen by first noting that, for $U$
and $V$ being invertible, \eqref{eq:MatrixEquation} is equivalent to
\begin{equation}
 U\cdot A\cdot q ~=~D\cdot V^{-1}\cdot q ~=~0 \mod N\;.
\end{equation}
This implies that there exists a linear combination of charges with integer
coefficients that sum up to $d_{n_q}=N$.  
We see immediately that $\rank(D\mod N) = \dim(A)-1$.

The constraint equation is now  brought to the diagonal form in the ``charge
basis'' $q' = V^{-1} \cdot q$ that are linear combinations of $q_{i}$'s
with integer coefficients,  
\begin{equation}\label{eq:Vq2}
 V^{-1}\cdot q~=~ \left(\begin{array}{c}
 k_1\,\frac{d_{n_q}}{d_1} \\ \vdots \\ k_{n_q}\,\frac{d_{n_q}}{d_{n_q}}
 \end{array}\right)  \;.
\end{equation}
The possibly inequivalent charges are thus given by
\begin{equation}\label{eq:Vq}
 q~=~V\cdot\left(\begin{array}{c}
 k_1\,\frac{d_{n_q}}{d_1} \\ \vdots \\ k_{n_q}\,\frac{d_{n_q}}{d_{n_q}}
 \end{array}\right) 
 \;.
\end{equation}
If we shift the charges $q_{i}$ by integral multiples of $N$, the r.h.s.\ of
\eqref{eq:Vq2} will shift by integral multiples of
$\gcd\bigl(\left(V^{-1}\right)_{i1},\dots \left(V^{-1}\right)_{in_q}\bigr)\cdot
N$ with $N=d_{n_q}$. Such shifts of the charges will obviously lead to the same
\Z{N} symmetry. However, since $V^{-1}$ is unimodular,
$\gcd\bigl(\left(V^{-1}\right)_{i1},\dots \left(V^{-1}\right)_{in_q}\bigr)=1$
for all $i$,  and we can take $k_i$ to lie only between 1 and $d_i$. We thus
obtain the charges for the maximal meaningful symmetry
$\Z{N=d_{n_q}}^{(R)}$ with the desired properties.  

We note that there exist more symmetries that fulfill the conditions.
Specifically, these additional symmetries can be obtained by dividing the order
$d_{n_q}$ by one of its divisors $\delta_i$. Then, \eqref{eq:MatrixEquation} will
still be fulfilled modulo $N'=d_{n_q}/\delta_i$. However, not all of these
symmetries possess all of the properties that $\Z{N}$ might have.

If there are inequalities, such as
\begin{equation}
3 q_Q + q_L - 2 q_{\theta} ~=~p~\ne~0\mod{N} \; ,
\end{equation}
all one has to do is to add
\begin{equation}
3 q_Q + q_L - 2 q_{\theta} - p~=~0\mod{N} 
\end{equation}
to the equation, regard $p$ as an extra charge, and project on solutions which
give $p\ne0\pmod N$. This will not lead to any new constraints. Therefore, one
can just scan the symmetries obtained from the imposed  equalities and explore
the possible $p$ values.

One can also determine the order $M$ in the inhomogeneous equation
\begin{equation}\label{eq:inhom1}
 A\cdot q~=~b\mod M
\end{equation}
with some $n_c$--dimensional vector $b$. After bringing $A$ to Smith normal form
and multiplying \Eqref{eq:inhom1} with $U$ from the left, we obtain
\begin{equation}
 D\cdot q'~=~b'\mod M
\end{equation}
with
\begin{equation}
 b'~=~U\cdot b
 \qquad\text{and}\qquad
 q'~=~V^{-1} \cdot q\;.
\end{equation}
If $n_c>n_q$, $b'$ can have non--trivial entries at the positions $n_q+1,\dots
,n_c$. Then, a solution is only possible if $M$ divides $b'_{n_q+1},\dots
b'_{n_c}$. Hence, we see that the maximal meaningful order may even be even more
constrained for inhomogeneous equations. An application of our methods will be
discussed in \Secref{sec:Z3R}.

In conclusion, we have looked at symmetries that fulfill certain constraint
equations. We have focussed on systems in which the constraint equations do not
allow for continuous or \U1 solutions. We have then shown that the maximal
meaningful order of \Z{N} symmetries compatible with the constraints can be read off
from the Smith normal form \eqref{eq:SmithForm} of the matrix encoding the
constraint equations, and is given by the last diagonal element $d_{n_q}$ (cf.\
equation~\eqref{eq:MatrixD}).

\subsection{Anomaly (non--)universality}
\label{sec:AnomalyUniversality}

As mentioned, anomalies for discrete symmetries can be cancelled by a discrete
version of the Green--Schwarz (GS) mechanism \cite{Green:1984sg}. This,
however, may destroy the beautiful picture of the MSSM gauge coupling
unification if the anomalies are not universal, i.e.\ if the GS axion couples
with different coefficients to the various field strength terms of the SM gauge
group factors.

We start out by discussing anomaly (non--)universality.
For a $\U1_{(R)}$ symmetry, the relevant anomaly coefficients are
\begin{subequations}\label{eq:AnomalyCoefficients}
\begin{align}
	A_3
	&~=~ \frac{1}{2}\sum\limits_{f}
	\Bigl[2 q_{Q}^{f} + q_{\overline{U}}^{f}+q_{\overline{D}}^{f}
	- 4 q_{\theta}\Bigr] + 3 q_{\theta}
    \notag \\
	&~=~\frac{3}{2}
	\Bigl[2 q_{Q} + q_{\overline{U}}+q_{\overline{D}}\Bigr] -3 q_{\theta}
	\;,\\
	A_2
	&~=~ \frac{1}{2}
	\Bigl[q_{\Hu} + q_{\Hd} - 2 q_{\theta} +
	\sum\limits_{f} \Bigl( 3 q_{Q}^{f} + q_{L}^{f} - 4 q_{\theta}  \Bigr) \Bigr]
   + 2 q_{\theta}\notag\\
   &~=~
   \frac{1}{2}
	\Bigl[q_{\Hu} + q_{\Hd} +
	3\,\Big( 3 q_{Q} + q_{L}  \Big) \Bigr]
   -5 q_{\theta}	
	\;,\\
	A_1
	&~=~ \frac{1}{2}\, \Bigl[ q_{\Hu} + q_{\Hd}
	- 2 q_{\theta} + \frac{1}{3} \sum\limits_{i} \Bigr( q_{Q}^{f} + 8
	q_{\overline{U}}^{f}+2 q_{\overline{D}}^{f} + 3 q_{L}^{f}
+ 6 q_{\overline{E}}^{f} - 20 q_{\theta} \Bigr)
									 \Bigr]\, Y_L^2  \notag\\
	&~=~ \frac{3}{10}\, \Bigl[ q_{\Hu} + q_{\Hd} + q_{Q} + 8
	q_{\overline{U}}+2 q_{\overline{D}} + 3 q_{L}
+ 6 q_{\overline{E}} - 22 q_{\theta} \Bigr]\;.
\end{align}
\end{subequations}
In the second line of each equation we switched to family--independent
charges. $q_\theta$ denotes the charge of the superspace coordinate $\theta$,
i.e.\ $q_\theta=0$ for a non--$R$ symmetry. $Y_L$ controls the normalization of
hypercharge, i.e.\ $Y_L^2=3/5$ if $\U1_Y$ is part of a unified \SU5 symmetry.

By imposing the existence of the Yukawa couplings we can eliminate the charges
of the $\overline{U}$, $\overline{D}$ and $\overline{E}$ fields,
\begin{subequations}
\label{eq:YukawaConstraints}
\begin{align}
 q_{\overline{U}}&~\equiv~-q_Q-q_{\Hu}+2q_\theta\;,\\
 q_{\overline{D}}&~\equiv~-q_Q-q_{\Hd}+2q_\theta\;,\\
 q_{\overline{E}}&~\equiv~-q_L-q_{\Hd}+2q_\theta\;.
\end{align}
\end{subequations}
where `$\equiv$' means `equal modulo $N$'. After eliminating $q_{\overline{U}}$,
$q_{\overline{D}}$ and $q_{\overline{E}}$ via \eqref{eq:YukawaConstraints}, the
anomaly coefficients \eqref{eq:AnomalyCoefficients} become
\begin{subequations}
\begin{align}
 A_3&~=~-\frac{3}{2}(q_{\Hu}+q_{\Hd}-2q_\theta)\;,\\
 A_2&~=~\frac{1}{2}\bigl[q_{\Hu}+q_{\Hd}+3\,(q_L+3q_Q)-10q_\theta\bigr]\;,\\
 A_1&~=~-\frac{3}{10}\bigl[7\,(q_{\Hu}+q_{\Hd})+3\,(q_L+3q_Q)-10q_\theta\bigr]\;.
\end{align}
\end{subequations}

By allowing for different couplings of the axion $a$ to the field strengths of
$\SU3_\mathrm{C}$, $\SU2_\mathrm{L}$ and $\U1_Y$, it is always possible to
cancel the anomalies  with the Green--Schwarz
mechanism\cite{Ludeling:2012cu,Dine:2012mf}. However, if one is to preserve
gauge coupling unification in a natural way, the anomalies need to be universal,
i.e.\
\begin{equation} \label{eq:AnomalyUniversality}
 A_3~=~A_2~=~A_1\;.
\end{equation}
For a discrete $\Z{N}^R$ symmetry, the coefficients in
\eqref{eq:AnomalyCoefficients} are only defined up to modulo
\begin{equation}
 \eta~=~\left\{\begin{array}{ll}
  N/2 & \text{if $N$ is even}\;,\\
  N & \text{if $N$ is odd}\;.
 \end{array}\right.
\end{equation}
The anomaly universality condition \eqref{eq:AnomalyUniversality} then boils
down to
\begin{equation} \label{eq:DiscreteAnomalyUniversality}
 A_3~\equiv~A_2~\equiv~A_1\;,
\end{equation}
where now `$\equiv$' means modulo~$\eta$.

Let us note that in \DHL\ the anomaly universality condition has been taken to
be $A_3\equiv A_2$. However, it is crucial to include the anomaly coefficient
due to $\U1_Y$,  the $A_{\U1_Y-\U1_Y-\Z{N}^R}$, particularly when addressing
compatibility with gauge coupling unification. Therefore, we will employ in
the first part of our analysis the universality condition
\eqref{eq:DiscreteAnomalyUniversality}.

The discrete anomaly universality conditions can be rewritten as
\begin{subequations}\label{eq:DiscreteAnomalyUniversality2}
\begin{align}
 A_3-A_2&~=~-2 q_{\Hd}-2 q_{\Hu}
 -\frac{3}{2}q_L -\frac{9}{2}q_Q+8q_\theta~\equiv~0\;,
 \label{eq:DiscreteAnomalyUniversality2a}\\
 A_3-A_1&~=~
 \frac{3}{10} (2 q_{\Hd}+2q_{\Hu}+3 q_L+9 q_Q)~\equiv~0\;.
 \label{eq:DiscreteAnomalyUniversality2b}
\end{align}
\end{subequations}
It is interesting to note that the second universality condition does not distinguish between
$R$ and non--$R$ symmetries, since it is independent of $q_\theta$.
By using the freedom of shifting $q_{\Hu}$ and $q_L$ by integral multiples of the
order $N$, we can shift the l.h.s.\ of \eqref{eq:DiscreteAnomalyUniversality2a}
by integral multiples of $N/2$ and the l.h.s.\ of
\eqref{eq:DiscreteAnomalyUniversality2b} by integral multiples of $3N/10$.
These equations then become equivalent to the so--called linear congruences
(see appendix~\ref{app:Congruence})
\begin{subequations}
\begin{align}
 N\,x&~=~2\,(A_3-A_2)\mod N\;,
 \label{eq:LineaerCongruence1}\\
 3N\,x&~=~10\,(A_3-A_1)
 \mod 10N\;.
 \label{eq:LineaerCongruence2}
\end{align}
\end{subequations}
 Since $\gcd(N,N)=\gcd(3N,10N)=N$, 
one obtains the constraints
\begin{subequations}\label{eq:AnomalyUniversality2}
\begin{align}
\label{eq:anom1}
 2\,(A_3-A_2)&
 ~=~
 -4 q_{\Hd}-4 q_{\Hu}
 -3q_L -9q_Q+16q_\theta
 ~\stackrel{!}{=}~0\mod N\;,\\
\label{eq:anom2}
 10\,(A_3-A_1)&
 ~=~
 3\,(2 q_{\Hd}+2q_{\Hu}+3 q_L+9 q_Q)
 ~\stackrel{!}{=}~0\mod N\;.
\end{align}
\end{subequations}
These constraints are now of the same type as the conditions for
operators in the superpotential or K\"ahler potential to be allowed.

In addition to the GS anomaly cancellation, we shall comment on
conditions for anomaly cancellation in the traditional sense, i.e.\
$A_3~\equiv~A_2~\equiv~A_1~\equiv~0$. This condition is equivalent
to demanding anomaly universality and $A_3~\equiv~0$. Consequently,
\begin{equation} \label{eq:RegularDiscreteAnomalyUniversality}
 2 A_3~=~-3 (q_{\Hu} + q_{\Hd} - 2 q_{\theta})~\stackrel{!}{=}~0\mod N\;.
\end{equation}
Thus, an anomaly--free symmetry can only forbid the $\mu$ term
under certain conditions. For $N=3$, the condition is trivially fulfilled.  If 3
divides $N$, then the solutions are given by $(q_{\Hu} + q_{\Hd} - 2
q_{\theta})\in\frac{N}{3}\cdot\{0,1,2\}$. As an example of the latter
case, consider an $\SU5$ compatible $\Z6^R$ symmetry of \cite{Lee:2011dya}. 
With the field charges of $(q_{\Hu}, q_{\Hd}, q_{\theta})~=~(4, 0, 1)$,
condition \eqref{eq:RegularDiscreteAnomalyUniversality} is satisfied
while the $\mu$ term is forbidden.

\subsection{Limitations of analysis}

While our aim is to provide a general analysis of discrete symmetries of the
MSSM with the properties discussed above, we note that our approach is not
without limitations. In particular, if there exist additional states at lower
energies, one can have effective operators which appear to have a total $R$
charge different from the one of the superpotential and which are endowed with
large coefficients. 

As an example, consider the MSSM with a $\Z4^R$ symmetry in which the dangerous
operator  $\overline{U}\,\overline{U}\,\overline{D}\,\overline{E}$ arise in the
K\"ahler potential with a  highly suppressed coefficient of $m_{3/2} /
M_{\mathrm{P}}^{3}$. This is based on the model which will be  specified in
Table~\ref{tab:Z4RchargesGSRPC} and assumes that the operator results from
integrating out heavy state(s) in the UV theory. 

On the other hand, adding a color triplet $X$ and an anti--triplet $\overline{X}$, both with
$R$ charge 0, we obtain additional allowed terms 
\begin{equation}
 \Delta\mathscr{W}~=~m_{\nicefrac{3}{2}}\,X\,\overline{X}+
 \overline{D}\,\overline{E}\,X+\overline{U}\,\overline{U}\,\overline{X},
\end{equation}
where we omitted the coefficients. After integrating out $X$ and $\overline{X}$
we get
\begin{equation}\label{eq:DeltaWeff1}
 \Delta\mathscr{W}_\mathrm{eff}
 ~=~
 \frac{1}{m_{\nicefrac{3}{2}}}\,
 \overline{U}\,\overline{U}\,\overline{D}\,\overline{E}\;,
\end{equation}
which is a dangerous proton decay operator with a large coefficient. 
 On the other hand, the
$\overline{U}\,\overline{U}\,\overline{D}\,\overline{E}$ operator has $R$ charge 0,
 i.e.\ according to our previous arguments we expect it to be suppressed.
To clarify this point, we note that  this operator is still $\Z4^R$
covariant since we can write 
\eqref{eq:DeltaWeff1} as~\cite{Lee:2011dya,Lee:2010gv} 
\begin{equation}
 \Delta\mathscr{W}_\mathrm{eff}
 ~=~
 \mathrm{e}^{\beta\,S}\,
 \overline{U}\,\overline{U}\,\overline{D}\,\overline{E}\;,
\end{equation}
where $S$ is the superfield that contains the axion and $\beta$ is a
coefficient. Note that this effective term has the opposite sign in the
exponential than the usual instanton contributions.   

Our analysis thus only applies if there are no extra states below the
fundamental scale $\Lambda$. Similar conclusions arise in the recently proposed
models of ``dynamical $R$ parity violation'' \cite{Csaki:2013jza}. 


\section{Models}
\label{sec:models}

\subsection{Examples of maximal meaningful symmetries }
\label{sec:max_sym}

First, as a cross--check of the algorithm, we have confirmed the maximal
meaningful order of a $\Z{N}^{R}$ symmetry in the MSSM that allows Yukawa
couplings and the Weinberg operator $\mathcal{O}_5$, which is in agreement with
previous analyses. Specifically, the maximal meaningful order
\begin{itemize}
 \item[(i)] for matter field charges satisfying \SU5 relations is 24 \cite{Lee:2011dya,Chen:2012jg};
 \item[(ii)] for matter field charges satisfying \SO{10} relations is 4 \cite{Lee:2010gv,Chen:2012jg}.
\end{itemize}
An explicit example for how the algorithm works can be found
in Appendix~\ref{app:ExampleZNsearch}.

As another illustration, we  consider symmetries compatible with Pati--Salam
partial unification. We demand the existence of the Yukawa couplings as well as
the Weinberg operator. Starting with the Pati--Salam charge relations
\begin{equation}
 q_Q~=~q_L
 \quad\text{and}\quad
 q_{\overline{U}}~=~q_{\overline{D}}~=~q_{\overline{E}}~,
\end{equation}
the maximal symmetry order is found to be 60. One of the inequivalent charge
assignments for a $\Z{60}^R$ symmetry is given by
\begin{subequations}
\begin{align}
 q_{\theta}&~=~1\;,\quad
 &q_{\Hu}&~=~q_{\Hd}~=~59~\equiv~-1\;,\\
 q_Q&~=~q_L~=~2
 \;,
 &q_{\overline{U}}&~=~q_{\overline{D}}~=~q_{\overline{E}}~=~1\;.
\end{align}
\end{subequations}
The $\mu$ term is forbidden; however, unlike in the case of the \SO{10} and
\SU5 compatible symmetries discussed above, it does not appear at linear order
in $m_{\nicefrac{3}{2}}$.

\subsection{\boldmath Pati--Salam compatible settings \unboldmath}
\label{sec:patisalam}

In contrast to $\SU{5}$ and $\SO{10}$, the Pati--Salam (PS) partial unification
\cite{Pati:1974yy} can be reconciled more easily with RPV. We note that the
Pati--Salam group evades the no--go theorems for $R$ symmetries in
four--dimensional GUT models~\cite{Fallbacher:2011xg}. RPV models with an
underlying PS symmetry have not been extensively studied, a gap which we aim to fill.

Specifically, we consider 4D Pati--Salam models with
$G_\mathrm{PS}=\SU4\times\SU2_\mathrm{L}\times\SU2_\mathrm{R}$ spontaneously
broken to $G_\mathrm{SM}$ by  the VEV of a $D$--flat combination of 
$\left(\rep{4},\rep{1},\rep{2}\right)\oplus\left(\crep{4},\rep{1},\rep{2}\right)$
Higgses with $R$ charge 0.  This VEV may then explain the effective  coupling
$\overline{U}\,\overline{D}\,\overline{D}$ or $L\,L\,\overline{E}$. In addition,
we would need Higgses in the $(\rep{6},\rep{1},\rep{1})$ and
$(\rep{1},\rep{1},\rep{1})$ representations with $R$ charge 2. Pati--Salam
models of this type have been derived from the heterotic string
\cite{Kobayashi:2004ya}.

We note that since the PS group does not fully unify into a single gauge group,
one can allow for different couplings of the GS axion to the different SM gauge
factors. In other words, PS does no lead to anomaly universality, which is
consistent with the fact that the PS symmetry does not imply gauge coupling
unification.

Let us now have a look at RPV models which are compatible with PS. As a first
example, we show that
\begin{equation}\left.\begin{array}{r}
\text{PS compatibility}\\
\text{allow $\overline{U}\,\overline{D}\,\overline{D}$}\\
\text{forbid $L\, \Hu$}\\
\end{array}\right\} \curvearrowright
\text{Weinberg operator is forbidden.}
\end{equation}
Starting with the PS compatibility, which implies
\begin{equation}
q_Q~=~q_L \;, ~q_{\overline{U}}~=~q_{\overline{D}}~=~q_{\overline{E}} \;,
~ \text{and}~~ q_{\Hu}~=~q_{\Hd}\;,
\end{equation}
one can now write down the conditions for the
$\overline{U}\,\overline{D}\,\overline{D}$ operator being allowed and the $L \,
\Hu$ term being forbidden,
\begin{align}
-3 q_{\Hu}-3q_L+4 q_\theta~&=~0 \mod{N} ~~~ (\overline{U}\,\overline{D}\,\overline{D})
\;, \\
q_{\Hu}+q_L-2 q_\theta~&\neq~0 \mod{N}~~~ (L \, \Hu)  \;.
\end{align}
Here we have taken into account the conditions for the existence of the
Yukawa couplings by the means of
\eqref{eq:YukawaConstraints}.
This leads to
\begin{equation}
2q_{\Hu}+2q_L-2 q_\theta~\neq~0 \mod{N} \;,
\end{equation}
which forbids the Weinberg operator. This result may be interpreted as the
statement that PS compatible $\cancel{B}$ RPV models tend to favor Dirac
neutrino masses.

\subsection{Scenarios with anomaly universality}
\label{sec:anom_univ}

\subsubsection{\boldmath Effective $R$ parity conservation
(RPC$_\mathrm{eff}$)\unboldmath}
\label{sec:rpc_anomuni}

We start by looking at scenarios which effectively preserve $R$ parity, in
which the usual $R$ parity violating operators are forbidden. However, we do not
explicitly impose $R$ parity.  For our search, we forbid dimension 4 and 5 RPV
operators in the superpotential, as well as the perturbative level $\mu$ term.
This leads to the following criteria
\begin{equation} 
 \text{RPC}_ \mathrm{eff} ~\curvearrowright~ \left\{ \begin{array}{rl}
2 q_{\Hd} + q_{\Hu} + 3 q_Q - 4 q_{\theta} &\neq~0 \mod{N} ~~~(\overline{U} \, \overline{D} \, \overline{D})
\;, \\
q_L - q_{\Hd} &\neq~0 \mod{N} ~~~(L \, L \, \overline{E})\;, \\
q_{\Hu} + q_{\Hd} - 2 q_{\theta} &\neq~0 \mod{N} ~~~(\Hu \, \Hd)\;, \\
q_L + q_{\Hu} - 2 q_{\theta} &\neq~0 \mod{N} ~~~(L\, \Hu)\;, \\
2 q_{\Hd} + 2 q_{\Hu} + q_L + 3 q_Q - 6 q_{\theta} &\neq~0 \mod{N}
~~~(\overline{U} \, \overline{U} \, \overline{D}\, \overline{E})\;, \\
3 q_Q + q_L - 2 q_{\theta}  &\neq~0 \mod{N} ~~~(Q \, Q\, Q\, L)\;, \\
3 q_Q + q_{\Hd} - 2 q_{\theta} &\neq~0 \mod{N} ~~~(Q\, Q\, Q \, \Hd)\;.
\end{array} \right.
\label{eq:pRPCuniv}
\end{equation}

With the various dimension 5 operators being related (see
\Secref{sec:strategy}), prohibiting the $Q\, Q\, Q \, \Hd$ term  also
automatically forbids $\mathcal{O}_{10}$. Similarly,  forbidding $L \, \Hu$
implies the absence of the operators $\mathcal{O}_{4}$,  $\mathcal{O}_{7}$,
$\mathcal{O}_{8}$ and $\mathcal{O}_{9}$.  Additionally, we 
will discuss whether a given
solution is compatible with the type I seesaw mechanism, i.e.\ if it satisfies
\begin{eqnarray}
2 q_L + 2 q_{\Hu} - 2 q_{\theta} &=~0 \mod{N} ~~~(L \, \Hu \, L\, \Hu)\;.
\end{eqnarray}

The minimal (GS) anomaly--universal solution which satisfies conditions
\eqref{eq:pRPCuniv} is a $\Z4^R$ symmetry whose charge assignment is specified
in \Tabref{tab:Z4charges} and Hilbert basis \cite{Kappl:2011vi} provided in
Appendix~\ref{sec:HilbertZ4R}. We see that this symmetry does indeed contain $R$
parity,  as there is a residual $\Z2$ after $R$ symmetry breaking.

\begin{table}[!h!]
\begin{center}
\begin{tabular}{|c|c|c|c|c|c|c|c|c|}
\hline
 Field & $\vphantom{\stackrel{!}{=}}Q$ & $\overline{U}$ & $\overline{D}$ & $L$ & $\overline{E}$
 & \Hu & \Hd & $\theta$\\
\hline
 $\Z4^R$ & 1 & 1 & 1 & 1 & 1 & 0 & 0 & 1\\
\hline
\end{tabular}
\end{center}
\caption{Anomaly--universal $R$ parity conserving symmetry $\Z4^R$.} \label{tab:Z4RchargesGSRPC}
\label{tab:Z4charges}
\end{table}

This $\Z4^R$ is nothing but the well--known $\Z4^R$
symmetry~\cite{Babu:2002tx},  which was found to be the unique anomaly--free
$\Z{N}^R$ solution that commutes with $\SO{10}$~\cite{Lee:2010gv,Lee:2011dya}.
We note, however, that \cite{Lee:2010gv} obtained this symmetry by imposing
criteria different from ours.\footnote{The analysis of \cite{Lee:2010gv}
imposed compatibility with $\SO{10}$, GS anomaly cancellation, absence of the
$\mu$ term before $R$ symmetry breaking, existence of the Yukawa couplings and
the presence of the Weinberg operator. On the other hand, we have obtained this
result by extending the analyses of \cite{Dreiner:2005rd, Ibanez:1991pr} to
allow for the Green--Schwarz mechanism, Yukawa couplings and by forbidding the
relevant dimension 4 and 5 RPV operators as well as the $\mu$ term in the
superpotential.}
The Weinberg operator as well as the Giudice--Masiero mechanism
\cite{Giudice:1988yz} for generating an effective $\mu$ term are both
compatible with this symmetry.

Interestingly, there exist solutions which ensure $R$ parity conservation before
SUSY breaking, but do not contain an actual $R$ parity. A $\Z{12}^R$ symmetry
with the charges given in \Tabref{tab:Z12RchargesGSRPCu} is of this type. The
$R$ parity violating operators get induced after the ``hidden sector"
superpotential acquires its VEV, and appear thus with coefficients that are
given by (high) powers of $m_{\nicefrac{3}{2}}/M_\mathrm{P}$. One thus
obtains a Froggatt--Nielsen--like \cite{Froggatt:1978nt} suppression of
these operators.

\begin{table}[!h!]
\begin{center}
\begin{tabular}{|c|c|c|c|c|c|c|c|c|}
\hline
 Field & $\vphantom{\stackrel{!}{=}}Q$ & $\overline{U}$ & $\overline{D}$ & $L$ & $\overline{E}$
 & \Hu & \Hd & $\theta$\\
\hline
 $\Z{12}^R$ & 4 & 4 & 0 & 0 & 4 & 6 & 10 & 1\\
\hline
\end{tabular}
\end{center}
\caption{Anomaly--universal effective $R$ parity conserving symmetry $\Z{12}^R$.} \label{tab:Z12RchargesGSRPCu}
\end{table}

\subsubsection{\boldmath $B$ violation at the renormalizable level \unboldmath}
\label{sec:BRPVanomuniv}

For the case of baryon number violating RPV setting, we impose the existence of
the $\overline{U} \, \overline{D} \, \overline{D}$ operator, and, at the
same time, the absence of the $L \, L \, \overline{E}$ term. Following \DHL,
the full set of phenomenological constraints can be specified as
\begin{equation}\label{eq:BRPV}
 \text{$\cancel{B}$ RPV} ~\curvearrowright~ \left\{ \begin{array}{rl}
2 q_{\Hd} + q_{\Hu} + 3 q_Q - 4 q_{\theta} &=~0 \mod{N} ~~~(\overline{U} \, \overline{D} \, \overline{D})\;, \\
q_L - q_{\Hd} &\neq~0 \mod{N} ~~~(L \, L \, \overline{E})\;, \\
q_{\Hu} + q_{\Hd} - 2 q_{\theta} &\neq~0 \mod{N} ~~~(\Hu \, \Hd)\;, \\
q_L + q_{\Hu} - 2 q_{\theta} &\neq~0 \mod{N} ~~~(L\, \Hu)\;, \\
3 q_Q + q_L - 2 q_{\theta}  &\neq~0 \mod{N} ~~~(Q \, Q\, Q\, L)\;. \\
\end{array} \right. 
\end{equation}
Additionally, we will require that the $L \, \Hu \, \Hd \, \Hu$ term be
absent.\footnote{If the $\cancel{L}$ operator $L \, \Hu \, \Hd \, \Hu$ is allowed,
its combination with the $\cancel{B}$ term $\overline{U} \, \overline{D} \,
\overline{D}$ could result in a fast proton decay. \DHL\ argue that, since the
relevant $\overline{U} \, \overline{D} \, \overline{D}$ coupling  contributing
to such process is $\lambda_{112}^{''}$, which is already strongly bounded by
the experiment \cite{Barbier:2004ez,Goity:1994dq,Allanach:1999ic}, the  $L \,
\Hu \, \Hd \, \Hu$ operator needs not be explicitly forbidden. However, we will
take on a more conservative position, and impose its absence.} 
This results in
\begin{eqnarray}  \label{eq:dangerop1}
q_L + q_{\Hd} + 2 q_{\Hu} - 2 q_{\theta} ~&\neq~0 \mod{N} ~~~(L \, \Hu \, \Hd \, \Hu)
\end{eqnarray}
A complete list of unique (GS) anomaly--universal solutions up to order 12,
satisfying the constraints of \eqref{eq:BRPV} and \eqref{eq:dangerop1}, can be
found in \Tabref{tab:BviolatingSymmetries} in Appendix \ref{app:brpv}. This set
contains a $\widetilde{\mathbbm{Z}}_8^R$ symmetry with the charge assignment of
\Tabref{tab:Z8RchargesGSB}. This $\widetilde{\mathbbm{Z}}_8^R$ symmetry is not
only compatible with the Pati--Salam group, but also allows for the
Giudice--Masiero mechanism to be implemented.

\begin{table}[!h!]
\begin{center}
\begin{tabular}{|c|c|c|c|c|c|c|c|c|}
\hline
 Field & $\vphantom{\stackrel{!}{=}}Q$ & $\overline{U}$ & $\overline{D}$ & $L$ & $\overline{E}$
 & \Hu & \Hd & $\theta$\\
\hline
 $\vphantom{\stackrel{!}{=}}\widetilde{\mathbbm{Z}}_8^R$ & 4 & 6 & 6 & 4 & 6 & 0 & 0 & 1\\
\hline
\end{tabular}
\end{center}
\caption{Anomaly--universal $\cancel{B}$ RPV symmetry $\widetilde{\mathbbm{Z}}_8^R$.} \label{tab:Z8RchargesGSB}
\end{table}

\subsubsection{\boldmath $L$ violation at the renormalizable level \unboldmath}
\label{sec:lviolanomuniv}

Similarly, we can identify (GS) anomaly--universal symmetries which violate
lepton number at the renormalizable level and satisfy the appropriate 
phenomenological constraints. However, a straightforward argument
\cite{Acharya:2014vha} appears to demonstrate that all such symmetries are
disfavored. Let us review this in more detail.

The argument of \cite{Acharya:2014vha} follows from the observation that if the
Yukawa couplings, the $\mu$ term as well as any of the trilinear leptonic RPV couplings
are allowed in the (perturbative) superpotential, so will be the $\kappa \,  L \, \Hu$ term. This leads
to the expectation that $\mu \sim \kappa$. In more detail,
equation 9 of \cite{Acharya:2014vha} states that
\begin{equation}
 q_L+q_{\Hu}
 ~=~
 q_Q+q_L+q_{\overline{D}}+q_\mu
 ~=~
 2q_L+q_{\overline{E}}+q_\mu\;,
\end{equation}
where $q_\mu=q_{\Hu}+q_{\Hd}-2q_\theta$ and the Yukawa conditions 
\eqref{eq:YukawaConstraints} have been used. This implies that, if the $\mu$
term is allowed, which implies that $q_\mu=0$, the charges of $L\,\Hu$,
$Q\,L\,\overline{D}$ and $L\,L\,\overline{E}$ coincide. Therefore, all these
symmetries are simultaneously allowed or forbidden by the symmetry. 
On the other hand, if we demand that $Q\,L\,\overline{D}$ and/or
$L\,L\,\overline{E}$ appear at the renormalizable level, i.e.\
$q_Q+q_L+q_{\overline{D}}=2q_\theta$ and/or $2q_L+q_{\overline{E}}=2q_\theta$,
then $q_L+q_{\Hu}=q_{\Hu}+q_{\Hd}$ such that one expects $\mu$ and $\kappa$ to
be of comparable orders.

Furthermore, suppose that the $\mu$ term originates from the K\"ahler
potential, while the trilinear leptonic RPV couplings are allowed in the
(perturbative) superpotential, i.e.\ before $R$ symmetry breaking, as in
the previous case. The same line of reasoning as above shows that $\kappa \,  L
\, \Hu$ will also be effectively generated with the size $\mu \sim \kappa$,
leading again to the same conclusion. Since the above argument applies to all
$\cancel{L}$ RPV models which have lepton number violating couplings present
before $R$ symmetry breaking, these scenarios are disfavored and we shall not
consider them further. Because neutrino mass generation from the bilinear
$\cancel{L}$ term is a popular mechanism in RPV settings (e.g.
\cite{Hall:1983id,Hirsch:2000ef,Mohapatra:2005wg,Grossman:2003gq}), the above
conclusion argument may be interpreted as a problematic feature on a
large class of models.

Let us comment, however, that the argument of \cite{Acharya:2014vha} is limited
in the following sense. The central assumption of the argument is that lepton
number violating couplings are present in the
superpotential before $R$ symmetry breaking. In contrast, if we require that
both the $\mu$ as well as the $\cancel{L}$ RPV terms arise only after $R$
symmetry breaking, the conclusion that any model with lepton number violation
must have  $\mu \sim \kappa$ can be evaded. We will demonstrate this with an
explicit example in \Secref{sec:nonuniv_nonpRPV},  where the operators arise
with different powers of $m_{\nicefrac{3}{2}}/M_\mathrm{P}$, thus leading to
very different sizes of $\mu$ and $\kappa$.

\subsection{Settings with anomaly non--universality}
\label{sec:anom_nonuniv}

As already mentioned, the Green--Schwarz anomaly cancellation may be satisfied
without requiring universality, if the GS axion couples differently to each MSSM
field strengths \cite{Ludeling:2012cu,Chen:2012tia}. Although this may spoil
precision gauge unification, dropping universality constraint leads to new
solutions. Such scenarios are not compatible with either \SU{5} or \SO{10}. 
However, as will be shown below, there exist solutions based on  the
Pati--Salam group with non--universal anomalies.

\subsubsection{\boldmath Effective $R$ parity conservation \unboldmath}

Abandoning anomaly universality, the lowest order solution which satisfies the
constraints of \eqref{eq:pRPCuniv} and is compatible with the Pati--Salam group
is a $\Z8^R$ symmetry with the charge given in \Tabref{tab:Z8RPCnonuniv}.
\begin{table}[!h]
\begin{center}
\begin{tabular}{|c|c|c|c|c|c|c|c|c|}
\hline
 field & $\vphantom{\stackrel{!}{=}}Q$ & $\overline{U}$ & $\overline{D}$ & $L$ & $\overline{E}$
 & \Hu & \Hd & $\theta$\\
\hline
 $\Z8^R$ & 1 & 1 & 1 & 1 & 1 & 0 & 0 & 1\\
\hline
\end{tabular}
\end{center}
\caption{Anomaly non--universal $R$ parity conserving symmetry $\Z8^R$.}\label{tab:Z8RPCnonuniv}
\end{table}
Clearly, after $R$ symmetry breaking, there is a residual \Z4 symmetry which
contains matter parity as a subgroup.

\subsubsection{\boldmath $B$ violation at renormalizable level  \unboldmath}

Imposing the phenomenological constraints of  \eqref{eq:BRPV} and
\eqref{eq:dangerop1} but allowing for anomaly coefficients to be non--universal,
the minimal solution with baryon number violation is a
$\widetilde{\mathbbm{Z}}_4^R$ symmetry with the charge assignment of
\Tabref{tab:Z4BRPVnonuniv}. This solution allows for the Giudice--Masiero
mechanism as well as the Weinberg operator. However, in contrast to the $\Z8^R$
symmetry in the anomaly--universal case discussed in 
\Secref{sec:BRPVanomuniv}, it does not commute with Pati--Salam.

\begin{table}[!htb!]
\begin{center}
\begin{tabular}{|c|c|c|c|c|c|c|c|c|}
\hline
 field & $\vphantom{\stackrel{!}{=}}Q$ & $\overline{U}$ & $\overline{D}$ & $L$ & $\overline{E}$
 & \Hu & \Hd & $\theta$\\
\hline
 $\vphantom{\stackrel{!}{=}}\widetilde{\mathbbm{Z}}_4^R$ & 0  & 2 & 2 & 1 & 1 & 0 & 0 & 1\\
\hline
\end{tabular}
\end{center}
\caption{Anomaly non--universal $\cancel{B}$ RPV symmetry $\widetilde{\mathbbm{Z}}_4^R$.}\label{tab:Z4BRPVnonuniv}
\end{table}

\subsubsection{\boldmath  ``Non--perturbative'' $B$ and $L$ violation\unboldmath}
\label{sec:Z3R}
\label{sec:nonuniv_nonpRPV}

Another interesting scenario is a setup where $R$ parity violation appears
after $R$ symmetry breaking. However, as we shall prove, there are no
phenomenologically viable anomaly--universal ``non--perturbative" RPV
symmetries.

Let us impose anomaly universality \eqref{eq:AnomalyUniversality2}  as well as
``non--perturbative" $B$ and $L$ violation. The latter condition amounts to the
requirement that the total charges of  $L \, L \, \overline{E}$ and
$\overline{U} \, \overline{D} \, \overline{D}$ are $0$. Since the number of
variables is larger than the number of independent equations, we can impose a
\U1 symmetry to satisfy all the constraints. Using the homogeneous version of the
method developed in \Secref{sec:SystematicSearch4ZN}, we obtain a single \U1
solution which has $q_\theta~=~0$, a continuous non--$R$ symmetry. However, as
pointed out in \cite{Chen:2012jg}, to forbid the $\mu$ term, one needs an $R$
symmetry with $q_\theta \neq 0$. Thus, we will look for discrete $R$ symmetries,
subject to the constraints
\begin{subequations}\label{eq:nonpRPVcon}
\begin{align}
 -4 q_{\Hd}-4 q_{\Hu}-3 q_L-9 q_Q+16 q_\theta ~&=~0 \mod{N}
\;, \\
 6 q_{\Hd}+6 q_{\Hu}+9\,\left(q_L+3 q_Q\right) ~&=~0\mod{N}
\;, \\
 -q_{\Hd}+q_L+2 q_\theta ~&=~ 0\mod{N}~~~(L \, L \, \overline{E})\;,\\
 -2 q_{\Hd}-q_{\Hu}-3q_Q+6 q_\theta ~&=~ 0\mod{N}~~~(\overline{U} \, \overline{D} \, \overline{D})\;.
\end{align}
\end{subequations}
Instead of treating $q_\theta$ as an extra variable, we can treat it as a
constant, and thus make use of the inhomogeneous variant of the method of
\Secref{sec:SystematicSearch4ZN}. Thus, one can rewrite \eqref{eq:nonpRPVcon} as
\begin{equation}\label{eq:Aq=b}
 A\cdot q~=~b\mod N
\end{equation}
with
\begin{equation}
A~=~
 \left(
\begin{array}{cccc}
 -4 & -4 & -3 & -9 \\
 6 & 6 & 9 & 27 \\
 -1 & 0 & 1 & 0 \\
 -2 & -1 & 0 & -3 \\
\end{array}
\right)\;,\quad
q~=~\left(\begin{array}{c}q_{\Hd}\\
q_{\Hu}\\
q_{L}\\
q_{Q}\end{array}\right)
\quad\text{and}\quad
b~=~q_\theta\cdot\left(\begin{array}{c}-16 \\ 0\\ -2  \\ -6 \end{array}\right)\;.
\end{equation}
Bringing $A$ to the Smith normal form, $D=U\cdot A \cdot V$, we can rewrite
\eqref{eq:Aq=b} as
\begin{equation}
 D\,q'~=~b'\mod N
 \qquad\text{with}
 \quad q'~=~V^{-1}\,q\quad\text{and}\quad b'~=~U\,b\;,
\end{equation}
where 
\begin{equation}
 D~=~\diag(1,1,1,0)
 \quad\text{and}\quad b'~=~(2 q_\theta ,2 q_\theta ,0,-24 q_\theta)\;.
\end{equation}
Clearly, the last equation $0\cdot q_4'=-24 q_\theta \mod N$ has only a solution
if $N$ is a divisor of $24$. We therefore conclude that the order of a discrete
$R$ symmetry that is consistent with our constraints has to divide 24. The
charges are then subject to the constraints
\begin{subequations}
\begin{align}
q_{\Hd}-q_L&~=~2 q_\theta \mod N\;,\\
q_{\Hu}+2 q_L+3 q_Q&~=~2 q_\theta \mod N\;,\\
q_L+3 q_Q&~=~0 \mod N\;.
\end{align}
\end{subequations}
Subtracting the last equation from the next--to--last one, we see that 
\begin{equation}
q_{\Hu}+q_L~=~2 q_\theta \mod N\;.
\end{equation}
From this we see that all such symmetries allow for the $\kappa\,\Hu\,L$ term in
the superpotential, and are, therefore, phenomenologically not viable. We have
hence proved that phenomenologically viable, anomaly--universal discrete $R$
symmetries that give rise to ``non--perturbative'' $B$ and $L$ violation do not
exist.

On the other hand, abandoning the anomaly universality condition allows us to
construct such models. For instance, a simple $\Z3^R$ symmetry can give rise to
scenarios with ``non--perturbative'' $B$ and $L$ violation. The charge
assignment for $\Z3^R$ can be found in Table \ref{tab:Z3charges} and the Hilbert
basis in Appendix~\ref{sec:HilbertZ3R}. 
\begin{table}[!h!]
\begin{center}
\begin{tabular}{|c|c|c|c|c|c|c|c|c|}
\hline
 field & $\vphantom{\stackrel{!}{=}}Q$ & $\overline{U}$ & $\overline{D}$ & $L$ & $\overline{E}$
 & \Hu & \Hd & $\theta$\\
\hline
 $\Z3^R$ & 1 & 1 & 1 & 1 & 1 & 0 & 0 & 1\\
\hline
\end{tabular}
\end{center}
\caption{Anomaly non--universal ``non--perturbative'' $\cancel{B}$ and $\cancel{L}$
symmetry $\Z3^R$.}
\label{tab:Z3charges}
\end{table}
If we assume that the $\Z3^R$ breaking is controlled by the gravitino mass
$m_{\nicefrac{3}{2}}$, we obtain effective RPV operators of the form
\begin{equation}
 \mathscr{W}_\mathrm{eff}^{\mathrm{RPV}}~\supset~
 \frac{m_{\nicefrac{3}{2}}}{M_\mathrm{P}}\,L\,L\,\overline{E}
 +
 \frac{m_{\nicefrac{3}{2}}}{M_\mathrm{P}}\,Q\,L\,\overline{D}
 +
 \frac{m_{\nicefrac{3}{2}}}{M_\mathrm{P}}\,\overline{U}\,\overline{D}\,\overline{D}
 \;.
\end{equation}
For $m_{\nicefrac{3}{2}}\sim\text{TeV}$, this implies that the couplings
$\lambda$, $\lambda'$ and $\lambda''$ are of the order $10^{-15}$. In addition,
there might exist further flavor suppression for the couplings of the lighter
generations. The $L\,\Hu$ term is suppressed by
$m_{\nicefrac{3}{2}}^2/M_\mathrm{P}$, while the $\mu$ term is of the order
$m_{\nicefrac{3}{2}}$. We have thus obtained a scenario with (sufficiently
suppressed) lepton number violation that provides, in some sense, a
counter--example to the statement in \cite{Acharya:2014vha} that in such
scenarios $\kappa\sim\mu$. Further, $Q\,Q\,Q\,L$ as well as
$\overline{U}\,\overline{U}\,\overline{D}\,\overline{E}$ are suppressed even
further by $m_{\nicefrac{3}{2}}^2/M_\mathrm{P}^3$.  Even though the charges in
this case commute with \SO{10}, precision gauge coupling unification might be
regarded as an accident in this setting,  as discussed in 
\Secref{sec:AnomalyUniversality}.

Even with $R$ parity preserved at the perturbative level, because of the
presence of the ``non--perturbative" RPV,  a sizable proton decay rate may still
exist.  Namely, the combination of $L \, Q \, \overline{D}$ and $\overline{U} \,
\overline{D} \, \overline{D}$ operators will lead to the usual $p \rightarrow
e^+ \pi^0$ proton decay channel. In our model, the relevant RPV couplings are
predicted to be $\lambda'~\sim~\lambda''~\sim~ 10^{-15}$, which leads to an
estimate on their combined strength $\lambda' \cdot \lambda''$ to be of order of
$10^{-30}$. This is value is to be compared to the experimental limit of
$\lambda' \cdot \lambda'' \lesssim 10^{-27}$ \cite{Raby:2012rw}.

We thus have provide a simple symmetry that gives rise to hierarchically small
$R$ parity violation. The LSP will be unstable. However, the gravitino will
still be a good dark matter candidate, as its decay rate will go like
$m_{\nicefrac{3}{2}}^5/M_\mathrm{P}^4$, where the Planck suppression originates
both from the fact that the gravitino interacts only gravitationally and that
the various $\lambda$ couplings go like $m_{\nicefrac{3}{2}}/M_\mathrm{P}$. A
complete survey of the phenomenology of this scenario is beyond the scope of the
present analysis.


\section{Summary}
\label{sec:summary}

The huge ratio between the GUT and electroweak scales allows us to give
compelling arguments for the observed longevity of the nucleon which is somewhat
hard to understand in extensions of the SM with low cut--off, where
higher--dimensional baryon and lepton number violating operators are not very
much suppressed. The traditional approach in supersymmetric model building is to
invoke matter or $R$ parity \cite{Fayet:1977yc}, amended by baryon triality
\cite{Ibanez:1991pr,Dreiner:2005rd}. More recently, a $\Z4^R$ symmetry
\cite{Babu:2002tx,Lee:2010gv}, which also solves the $\mu$ problem
\cite{Lee:2010gv,Lee:2011dya}, has been proposed. These symmetries remain the
simplest options to explain the longevity of the proton in supersymmetric
extensions of the SM.

On the other hand, nature might have taken a different route, and the $B$ or $L$
symmetries may be violated. In this study, we have explored discrete $R$
symmetries which explain a sufficient suppression of nucleon decay operators.
In settings with such symmetries, $R$ parity violation is related to
supersymmetry breaking, i.e.\ RPV couplings are suppressed by appropriate powers
of $m_{\nicefrac{3}{2}}/\Lambda$. In the course of our work, we completed  and
extended the analysis of \DHL\ surveying viable RPV symmetries of  the MSSM. 
We found that in some cases the symmetries are incompatible with the
Weinberg operator such that Dirac neutrino masses appear to be favored.
We allowed for $q_{\theta} \in \mathbbm{N}$ as
well as symmetries with non--universal anomalies. We identified redundant
symmetries in \DHL\ and found some new solutions. The most ``appealing''
solution that emerges for a given set of assumptions is depicted in Figures
\ref{fig:sol_part1} and \ref{fig:sol_part2}. Further, we have developed a novel
algorithm based on the Smith normal form, which allows to identify the maximal
meaningful symmetry for a given set of constraints. We also specified the
conditions for a given set of symmetries to be equivalent.  

\enlargethispage{0.5cm}
We have also identified a simple $\Z3^R$ symmetry that ensures $R$ parity
conservation before supersymmetry breaking. The coefficients of the $R$
parity violating operators are consequently suppressed by the small ratio
$m_{\nicefrac{3}{2}}/M_\mathrm{P}$. This symmetry provides us, in some
sense, with a counter--example to the statement in the recent literature 
that in lepton number violating scenarios, the $\mu$ term
and the bilinear $\kappa \, L \, \Hu$ must be of comparable magnitude.

\subsection*{Acknowledgments}

We would like to thank Csaba Cs\'aki for useful discussions. M.--C.C.\ would
like to thank TU M\"unchen, where part of the work  was done, for hospitality.
M.R.\ would like to thank the  UC Irvine, where part of this work was done, for 
hospitality. This work was partially supported by the DFG cluster  of excellence
``Origin and Structure of the Universe'' by Deutsche Forschungsgemeinschaft
(DFG) and by the Munich Institute for Astro- and Particle Physics (MIAPP).
The work of M.--C.C.\ was supported, in part, by the U.S.\ National Science 
Foundation under Grant No.\ PHY--0970173.  The work of V.T.\ was supported, in
part,  by the U.S.\ Department of Energy (DOE) under Grant No.\ DE--SC0009920. 
This work was supported in part by National Science Foundation Grant  No.\
PHYS--1066293  and NSF PHY11--25915. M.--C.C.\ and M.R.\ would like to thank the
hospitality of the Aspen Center for Physics.  M.--C.C.\ would like to thank the
Kavli Institute for Theoretical Physics at UCSB for hospitality.  This research
was done in the context of the ERC Advanced Grant project ``FLAVOUR''~(267104).
The project was supported by BaCaTec under the project number 6020039.

\tikzset{every node/.style={on grid=true,rounded corners,align=center,drop shadow},
    block/.style={rectangle,draw,top color= white,bottom color=blue!15,minimum height=4em,text width=3cm},
    yn/.style={rectangle,midway,fill=yellow!15,draw,text width=0.54cm,text depth=0pt},
    decision/.style={regular polygon,regular polygon sides=8,inner sep=0pt,draw,top color= white,bottom color=red!15},
    arrowFix1/.style={shorten <=-2pt},
    arrowFix2/.style={shorten >=-2pt},}

\begin{figure}[H]
\begin{center}
  \begin{tikzpicture}[thick,node distance=5.5cm,>=stealth]
    \node[decision] (ci) at (0,0) {\begin{tabular}{c}
$R$ parity\\
conserved\end{tabular}};
    \node[decision,right=5em of ci.east] (lambda) {\begin{tabular}{c}
$\lambda~\&~\lambda'\ne0$\\
	before\\ $\cancel{\text{SUSY}}$\end{tabular}};
    \node[decision,right=7em of lambda.east] (lambdaprime) {\begin{tabular}{c}
$\lambda''\ne0$\\
	before\\ $\cancel{\text{SUSY}}$\end{tabular}};
    \node[block,below=7em of ci.south] (Rparity) {$\Z4^R$ };
    \node[block,right=3em of Rparity.east] (lambda0) {$ - $};
    \node[block,right=3em of lambda0.east] (lambdaprime0) {$\widetilde{\mathbbm{Z}}_8^R$};
    \node[block,right=3em of lambdaprime0.east] (lambdaprimene0) {$\lambda\cdot\lambda''$ and $\lambda'\cdot\lambda''$ \\
small: $ - $};
    \draw[->,ultra thick] (ci.south) -- (Rparity.north) node[yn] {yes};
    \draw[->,ultra thick] (ci) -- (lambda) node[yn] {no};
    \draw[->,ultra thick] (lambda.east) -- (lambdaprime) node[yn,yshift=0.25em] {no};
    \draw[->,ultra thick] (lambda.south)-- (lambda0) node[yn] {yes};
    \draw[->,ultra thick] (lambdaprime.south west)|-+(0,-2.5em)-| (lambdaprime0) node[yn] {yes};
    \draw[->,ultra thick] (lambdaprime.south east)|-+(0,-2.5em)-| (lambdaprimene0)
	node[yn] {no};
  \end{tikzpicture}
\end{center}
\caption{Summary of our results. We present the simplest discrete $R$ symmetries
with universal anomalies and the specified properties. The symbol ``--''
indicates the absence of a solution.
}
\label{fig:sol_part1}
\end{figure}

\begin{figure}[H]
\begin{center}
  \begin{tikzpicture}[thick,node distance=5.5cm,>=stealth]
    \node[decision] (ci) at (0,0) {\begin{tabular}{c}
$R$ parity\\
conserved\end{tabular}};
    \node[decision,right=5em of ci.east] (lambda) {\begin{tabular}{c}
$\lambda~\&~\lambda'\ne0$\\
	before\\ $\cancel{\text{SUSY}}$\end{tabular}};
    \node[decision,right=7em of lambda.east] (lambdaprime) {\begin{tabular}{c}
$\lambda''\ne0$\\
	before\\ $\cancel{\text{SUSY}}$\end{tabular}};
    \node[block,below=7em of ci.south] (Rparity) {$\Z8^R$};
    \node[block,right=3em of Rparity.east] (lambda0) {$ - $};
    \node[block,right=3em of lambda0.east] (lambdaprime0) {$\widetilde{\mathbbm{Z}}_4^R$};
    \node[block,right=3em of lambdaprime0.east] (lambdaprimene0) {$\lambda\cdot\lambda''$ and $\lambda'\cdot\lambda''$ \\
small: $\Z3^R$};
    \draw[->,ultra thick] (ci.south) -- (Rparity.north) node[yn] {yes};
    \draw[->,ultra thick] (ci) -- (lambda) node[yn] {no};
    \draw[->,ultra thick] (lambda.east) -- (lambdaprime) node[yn,yshift=0.25em] {no};
    \draw[->,ultra thick] (lambda.south)-- (lambda0) node[yn] {yes};
    \draw[->,ultra thick] (lambdaprime.south west)|-+(0,-2.5em)-| (lambdaprime0) node[yn] {yes};
    \draw[->,ultra thick] (lambdaprime.south east)|-+(0,-2.5em)-| (lambdaprimene0)
	node[yn] {no};
  \end{tikzpicture}
\end{center}
\caption{Summary of our results. We present the simplest discrete $R$ symmetries
with non--universal anomalies and the specified properties. The symbol ``--'' indicates the absence of a solution.
}
\label{fig:sol_part2}
\end{figure}


\appendix

\section{Basic facts on congruences}
\label{app:Congruence}

In general, the linear congruence
\begin{equation}\label{eq:LinearCongruence}
 a\,x~=~b\mod M
\end{equation}
has solutions if and only if $b$ is divisible by $\gcd(M,a)$, in which case
there are $\gcd(M,a)$ solutions modulo $M$. Further, it is true that if
\begin{equation}
 a~=~ b\mod N \quad\text{and}\quad
 c~= ~d\mod N
\end{equation}
then
\begin{subequations}
\begin{align}
 a+c&~=~ b+d\mod N\;,\\
 a\cdot c &~=~ b \cdot d\mod N\;.
\end{align}
\end{subequations}

\section{\boldmath Example of systematic search for $\Z{N}^{(R)}$ \unboldmath}
\label{app:ExampleZNsearch}

In this example, we will impose anomaly universality, existence of
the $\overline{U}\,\overline{D}\,\overline{D}$ term, as well as  Pati--Salam 
compatibility $q_L=q_Q$ and the GM
condition $q_{\Hu}+q_{\Hd}=0\mod N$. After imposing these conditions, we are
left with the two charges  $\{q_i\}=\left\{q_Q,q_{\theta }\right\}$.

The conditions  are then encoded in the matrix equation
\begin{equation}
A\cdot q~=~0\mod N \qquad \text{with} \quad
A~=~
\left(
\begin{array}{cc}
 12 & -16 \\
 0 & 48 \\
 -6 & 8
\end{array}
\right)\;.
\end{equation}
The Smith normal form of $A$ is then given by the matrices
\begin{equation}
D~=~
\left(
\begin{array}{cc}
 2 & 0 \\
 0 & 144 \\
 0 & 0 \\
\end{array}
\right)\;,\quad
U~=~ \left(
\begin{array}{ccc}
 0 & 0 & -1 \\
 0 & 1 & -24 \\
 1 & 0 & 2 \\
\end{array}
\right)
\quad\text{and}\quad V~=~ \left(
\begin{array}{cc}
 -1 & 4 \\
 -1 & 3 \\
\end{array}
\right)\;.
\end{equation}
We see that the maximal meaningful $\Z{N}$ symmetry has $N=144$. The
corresponding charges are given by
\begin{subequations}
\begin{align}
 q_Q&~=~-72\cdot k_1+4\cdot k_2 \mod 144\;,\\
 q_\theta&~=~-72\cdot k_1+3\cdot k_2 \mod 144\;,
\end{align}
\end{subequations}
where $k_1\in\{1,2\}$ and $k_2\in\{1,\dots144\}$. However, as discussed in
\Secref{sec:EquivalentDiscreteSymmetries}, many different $k_i$
lead to physically equivalent symmetries. 
The resulting inequivalent symmetries, with the $\mu$ term forbidden,
are shown in \Tabref{tab:UDD}.

\clearpage

\vspace*{-1.5cm}

\begin{table}[H]
\begin{center}
\begin{tabular}{|ccccccccc|}
\hline
$N$ & $q_Q$ & $q_{\bar{U}}$ & $q_{\bar{D}}$ & $q_L$ & $q_{\bar{E}}$ & $q_{\Hu}$ &
$q_{\Hd}$ & $q_{\theta}$ \\
\hline
  4 & 0 & 2 & 2 & 0 & 2 & 0 & 0 & 1\\
  8 & 4 & 6 & 6 & 4 & 6 & 0 & 0 & 1\\
  9 & 1 & 5 & 5 & 1 & 5 & 0 & 0 & 3\\
  12 & 4 & 2 & 2 & 4 & 2 & 0 & 0 & 3\\
  16 & 4 & 6 & 6 & 4 & 6 & 8 & 8 & 1\\
  18 & 1 & 14 & 14 & 1 & 14 & 9 & 9 & 3\\
  24 & 4 & 2 & 2 & 4 & 2 & 0 & 0 & 3\\
  36 & 4 & 2 & 2 & 4 & 2 & 0 & 0 & 3\\
  48 & 4 & 2 & 2 & 4 & 2 & 0 & 0 & 3\\
  72 & 4 & 2 & 2 & 4 & 2 & 0 & 0 & 3\\
  144 & 4 & 2 & 2 & 4 & 2 & 0 & 0 & 3\\
\hline
\end{tabular}
\end{center}
\caption{$\Z{N}^R$ symmetries with renormalizable
$\overline{U}\,\overline{D}\,\overline{D}$, matter charges that commute with
PS and the Higgs charges 
which fulfill the GM condition $\qHu+\qHd=0\mod N$.}
\label{tab:UDD}
\end{table}

\section{\boldmath $\Z{N}^{(R)}$ symmetries of $B$ violating settings \unboldmath}
\label{app:brpv}

Here we list the $\Z{N\le12}^{(R)}$ inequivalent symmetries of settings with
renormalizable $\cancel{B}$. 

\begin{table}[H]
{\small
\[
\begin{array}{|ccccccccc|cccccccc|c|c|}
\multicolumn{9}{c}{\text{symmetry}} &
\multicolumn{8}{c}{\text{residual symmetry}}
& \multicolumn{2}{c}{} 
\\
\hline
N & Q & \overline{U} & \overline{D} & L & \overline{E} & \Hu & \Hd & \theta
&
N' & Q & \overline{U} & \overline{D} & L & \overline{E} & \Hu & \Hd
& W 
& \text{GS}\\
\hline
 5 & 2 & 2 & 0 & 2 & 0 & 3 & 0 & 1 & \multicolumn{8}{c|}{-} & - &  \checkmark \\
 6 & 1 & 2 & 5 & 1 & 5 & 3 & 0 & 0 & 6 & 1 & 2 & 5 & 1 & 5 & 3 & 0  & -  & \checkmark \\
 6 & 1 & 0 & 1 & 3 & 5 & 1 & 0 & 1 & 2 & 1 & 0 & 1 & 1 & 1 & 1 & 0 & \checkmark  &  \checkmark \\
 6 & 1 & 4 & 3 & 3 & 1 & 5 & 0 & 2 & 2 & 1 & 0 & 1 & 1 & 1 & 1 & 0  & \checkmark &  \checkmark \\
 8 & 4 & 6 & 6 & 4 & 6 & 0 & 0 & 1 & \multicolumn{8}{c|}{-} & -  & \checkmark \\
 9 & 1 & 2 & 8 & 1 & 8 & 6 & 0 & 0 & 9 & 1 & 2 & 8 & 1 & 8 & 6 & 0  & -  & - \\
 9 & 1 & 5 & 5 & 1 & 5 & 0 & 0 & 3 & 3 & 1 & 2 & 2 & 1 & 2 & 0 & 0 & -  & - \\
 10 & 2 & 2 & 0 & 2 & 0 & 8 & 0 & 1 & \multicolumn{8}{c|}{-} & -  & \checkmark \\
 10 & 7 & 2 & 5 & 7 & 5 & 3 & 0 & 1 & 2 & 1 & 0 & 1 & 1 & 1 & 1 & 0  & -  & \checkmark \\
 12 & 2 & 2 & 0 & 2 & 0 & 10 & 0 & 1 & \multicolumn{8}{c|}{-} & - & - \\
 12 & 0 & 10 & 2 & 4 & 10 & 4 & 0 & 1 & \multicolumn{8}{c|}{-}  & -  &  \checkmark \\
 12 & 0 & 10 & 2 & 8 & 6 & 4 & 0 & 1 & \multicolumn{8}{c|}{-} & - & \checkmark \\
 12 & 2 & 2 & 0 & 10 & 4 & 10 & 0 & 1 & \multicolumn{8}{c|}{-}  & - & - \\
 12 & 0 & 6 & 6 & 4 & 2 & 0 & 0 & 3 & 3 & 0 & 0 & 0 & 2 & 1 & 0 & 0   & - & \checkmark \\
\hline
\end{array}
\]
}
\caption{Anomaly--universal $\cancel{B}$ symmetries up to order 12. We specify
the residual symmetry after the breaking of the $R$ symmetry, and show in the W
column if the Weinberg operator $L\,\Hu\,L\,\Hu~(\mathcal{O}_5)$ is allowed. 
The last column indicates whether or not a non--trivial GS mechanism is at
work.}
\label{tab:BviolatingSymmetries}
\end{table}

\section{\boldmath RPV symmetries of DHL\unboldmath}
\label{app:full_dhl}

In what follows, we list and comment on the non--redundant symmetries found
by \DHL. We explicitly state whether these allow for a Giudice--Masiero
mechanism, i.e.\ if $q_{\Hu} + q_{\Hd}~=~0 \mod N$.
We further specify the order $N_{\mathscr{W}}'$ of the residual symmetry that
is left after the ``hidden sector" superpotential acquires a VEV.\footnote{Note
that in \DHL\ different residual symmetries are discussed. The charge of the
supersymmetry breaking spurion is adjusted in such a way that the $\mu$ term can
be generated through the Giudice--Masiero mechanism. However, in addition to the
breaking of the symmetry by the $F$ term VEV, there will always be the breaking
due to expectation value of the ``hidden sector" superpotential.}

\subsection{\boldmath $B$ violating settings\unboldmath}

\begin{table}[H]
 \[
 \centering
\footnotesize
 \begin{array}{|c|c|c|c|c|c|c|c|c|c|c|c|c|c|c|c|c|}
\hline
 N & \theta & p & n & m & N'_{\mathscr{W}} & 
\text{W} & \mathcal{O}_6 &
 \text{GM} & Q & \overline{U} & \overline{D}  & 
L &\overline{E}  & \Hu & \Hd &
\text{A} \\
\hline
 5 & 0 & 1 & 3 & 1 & 5  & - & \checkmark & - & 0 & 4 & 3 & 1 & 2 & 1 & 2  & - \\
 5 & 1 & 3 & 4 & 0 & -  & - & - & - & 0 & 0 & 1 & 3 & 3 & 2 & 1  & U \\
 6 & 0 & 1 & 3 & 0 & 6  & - & - & - & 0 & 0 & 3 & 2 & 1 & 0 & 3  & U \\
 6 & 1 & 3 & 3 & 2 & 2  & \checkmark & - & - & 0 & 4 & 5 & 0 & 5 & 4 & 3 & U \\
 7 & 0 & 1 & 3 & 6 & 7  & - & - & - & 0 & 1 & 3 & 3 & 0 & 6 & 4  & - \\
 7 & 1 & 1 & 3 & 1 & -  & - & - & \checkmark & 0 & 6 & 5 & 3 & 2 & 3 & 4  & - \\
 7 & 1 & 2 & 6 & 0 & -  & \checkmark & - & - & 0 & 0 & 1 & 6 & 2 & 2 & 1  & - \\
 7 & 1 & 5 & 1 & 4 & -  & - & - & - & 0 & 3 & 3 & 1 & 2 & 6 & 6  & - \\
 8 & 0 & 1 & 3 & 6 & 8  & - & - & - & 0 & 2 & 3 & 4 & 7 & 6 & 5  & - \\
 8 & 1 & 1 & 3 & 0 & 2  & - & - & - & 0 & 0 & 5 & 4 & 1 & 2 & 5  & - \\
 8 & 1 & 4 & 4 & 2 & -  & - & - & \checkmark & 0 & 6 & 6 & 0 & 6 & 4 & 4  & U \\
 8 & 1 & 6 & 2 & 6 & -  & - & - & - & 0 & 2 & 4 & 0 & 4 & 0 & 6  & - \\
 9 & 0 & 1 & 3 & 6 & 9  & - & - & - & 0 & 3 & 3 & 5 & 7 & 6 & 6  & 0 \\
 9 & 1 & 1 & 3 & 8 & - & - & - & - & 0 & 1 & 5 & 5 & 0 & 1 & 6  & - \\
 9 & 1 & 2 & 6 & 5 & -  & - & - & - & 0 & 4 & 8 & 1 & 7 & 7 & 3  & - \\
 9 & 1 & 3 & 0 & 2 & - & \checkmark & - & - & 0 & 7 & 2 & 6 & 5 & 4 & 0  & - \\
 9 & 1 & 5 & 6 & 5 & -  & - & - & - & 0 & 4 & 8 & 7 & 1 & 7 & 3  & - \\
 9 & 1 & 6 & 0 & 2 & -  & - & \checkmark & - & 0 & 7 & 2 & 3 & 8 & 4 & 0  & - \\
 9 & 1 & 7 & 3 & 8 & - & - & - & - & 0 & 1 & 5 & 8 & 6 & 1 & 6  & - \\
 10 & 0 & 1 & 3 & 6 & 10 & - & - & - & 0 & 4 & 3 & 6 & 7 & 6 & 7  & - \\
 10 & 0 & 2 & 6 & 2 & 5  & - & \checkmark & - & 0 & 8 & 6 & 2 & 4 & 2 & 4  & - \\
 10 & 1 & 1 & 3 & 8 & 2& \checkmark & - & - & 0 & 2 & 5 & 6 & 9 & 0 & 7  & - \\
 10 & 1 & 3 & 9 & 0 & 2& - & - & - & 0 & 0 & 1 & 8 & 3 & 2 & 1  & U \\
 10 & 1 & 5 & 5 & 2 & 2  & - & - & - & 0 & 8 & 7 & 0 & 7 & 4 & 5  & - \\
 10 & 1 & 8 & 4 & 0 & -  & - & - & - & 0 & 0 & 6 & 8 & 8 & 2 & 6 & U \\
\hline
\end{array} \]
\caption{$\Z{N\le12}^{(R)}$ symmetries for $B$ violating settings. Compatibility
with the Weinberg operator (W) and $L\,\Hu\,\Hd\,\Hu$ ($\mathcal{O}_6$) are
indicated. We specify if anomalies (A) vanish (0), are GS universal $(U)$, 
or fulfill neither of these conditions ($-$).}
\label{tab:Bviolation}
\end{table}

\subsection{\boldmath $L$ violating settings\unboldmath}

\begin{table}[H]
 \[
   \centering
	\footnotesize
\begin{array}{|c|c|c|c|c|c|c|c|c|c|c|c|c|c|c|c|c|}
\hline
 N & \theta & p & n & m & N'_{\mathscr{W}} & 
\text{W} & \mathcal{O}_2 &
 \text{GM} & Q & \overline{U} & \overline{D}  & 
L &\overline{E}  & \Hu & \Hd & \text{A} \\
\hline
 3 & 1 & 2 & 0 & 1 & - & \checkmark & \checkmark & - & 0 & 2 & 1 & 1 & 0 & 0 & 1 & 0 \\
 4 & 1 & 1 & 3 & 1 & 2  & \checkmark & \checkmark & - & 0 & 3 & 2 & 0 & 2 & 3 & 0  & - \\
 5 & 0 & 1 & 3 & 2 & 5  & - & - & - & 0 & 3 & 4 & 1 & 3 & 2 & 1  & - \\
 5 & 1 & 3 & 4 & 3 & - & - & - & - & 0 & 2 & 4 & 3 & 1 & 0 & 3  & U \\
 5 & 1 & 4 & 2 & 0 & -  & \checkmark & \checkmark & - & 0 & 0 & 3 & 4 & 4 & 2 & 4  & - \\
 6 & 0 & 1 & 3 & 1 & 6  & \checkmark & - & - & 0 & 5 & 4 & 2 & 2 & 1 & 2  & U \\
 6 & 1 & 2 & 0 & 4 & - & \checkmark & \checkmark & - & 0 & 2 & 4 & 4 & 0 & 0 & 4  & 0 \\
 6 & 1 & 3 & 3 & 5 & 2  & \checkmark & - & - & 0 & 1 & 2 & 0 & 2 & 1 & 0  & U \\
 6 & 1 & 5 & 3 & 1 & 2  & \checkmark & \checkmark & - & 0 & 5 & 4 & 4 & 0 & 3 & 4  & U \\
 7 & 0 & 1 & 3 & 0 & 7  & - & - & - & 0 & 0 & 4 & 3 & 1 & 0 & 3  & - \\
 7 & 1 & 1 & 3 & 2 & -  & - & - & \checkmark & 0 & 5 & 6 & 3 & 3 & 4 & 3 & - \\
 7 & 1 & 2 & 6 & 2 & -  & - & - & - & 0 & 5 & 3 & 6 & 4 & 4 & 6  & - \\
 7 & 1 & 5 & 1 & 2 & -  & - & - & - & 0 & 5 & 1 & 1 & 0 & 4 & 1  & - \\
 7 & 1 & 6 & 4 & 2 & -  & \checkmark & \checkmark & - & 0 & 5 & 5 & 4 & 1 & 4 & 4 & U \\
 8 & 0 & 1 & 3 & 7 & 8  & - & - & - & 0 & 1 & 4 & 4 & 0 & 7 & 4  & - \\
 8 & 1 & 1 & 3 & 1 & 2  & - & - & - & 0 & 7 & 6 & 4 & 2 & 3 & 4  & - \\
 8 & 1 & 3 & 1 & 7 & 2  & \checkmark & \checkmark & - & 0 & 1 & 6 & 4 & 2 & 1 & 4  & - \\
 8 & 1 & 4 & 4 & 6 & -  & - & - & \checkmark & 0 & 2 & 2 & 0 & 2 & 0 & 0 & U \\
 8 & 1 & 6 & 2 & 4 & -  & - & - & - & 0 & 4 & 2 & 0 & 2 & 6 & 0  & - \\
 9 & 0 & 1 & 3 & 7 & 9  & - & - & - & 0 & 2 & 4 & 5 & 8 & 7 & 5 & 0 \\
 9 & 1 & 1 & 3 & 0 & - & - & - & - & 0 & 0 & 6 & 5 & 1 & 2 & 5  & - \\
 9 & 1 & 2 & 6 & 7 & -  & \checkmark & - & - & 0 & 2 & 1 & 1 & 0 & 0 & 1  & - \\
 9 & 1 & 3 & 0 & 5 & - & - & - & - & 0 & 4 & 5 & 6 & 8 & 7 & 6  & - \\
 9 & 1 & 5 & 6 & 1 & - & \checkmark & - & - & 0 & 8 & 4 & 7 & 6 & 3 & 7  & - \\
 9 & 1 & 6 & 0 & 8 & -  & - & - & - & 0 & 1 & 8 & 3 & 5 & 1 & 3  & - \\
 9 & 1 & 7 & 3 & 6 & - & - & - & - & 0 & 3 & 3 & 8 & 4 & 8 & 8  & - \\
 9 & 1 & 8 & 6 & 4 & -  & \checkmark & \checkmark & - & 0 & 5 & 7 & 4 & 3 & 6 & 4  & - \\
 10 & 0 & 1 & 3 & 7 & 10  & - & - & - & 0 & 3 & 4 & 6 & 8 & 7 & 6  & - \\
 10 & 1 & 1 & 3 & 9 & 2  & - & - & - & 0 & 1 & 6 & 6 & 0 & 1 & 6  & - \\
 10 & 1 & 3 & 9 & 3 & 2 & - & - & - & 0 & 7 & 4 & 8 & 6 & 5 & 8  & U \\
 10 & 1 & 4 & 2 & 0 & -  & \checkmark & \checkmark & - & 0 & 0 & 8 & 4 & 4 & 2 & 4  & - \\
 10 & 1 & 5 & 5 & 7 & 2  & - & - & - & 0 & 3 & 2 & 0 & 2 & 9 & 0  & - \\
 10 & 1 & 8 & 4 & 8 & -  & - & - & - & 0 & 2 & 4 & 8 & 6 & 0 & 8  & U \\
 10 & 1 & 9 & 7 & 5 & 2 & \checkmark & \checkmark & - & 0 & 5 & 8 & 4 & 4 & 7 & 4  & - \\
\hline
\end{array} \]
\caption{$\Z{N\le12}^{(R)}$ symmetries for $L$ violating settings. The columns
have the same meaning  as in table~\ref{tab:Bviolation}.}
\label{tab:Lviolation}
\end{table}

\section{\boldmath Hilbert bases \unboldmath}
\label{app:hilbert}

The Hilbert basis method \cite{Kappl:2011vi} allows us to construct a complete
basis for the gauge invariant monomials $\mathscr{M}_i$ of fields appearing in the superpotential. In the
presence of $R$ symmetries, the $\mathscr{M}_i$ decompose into homogeneous and
inhomogeneous monomials, where allowed superpotential terms are of the form
\begin{equation}
 \mathscr{W}~\supset~
 \mathscr{M}_\mathrm{inhom}\cdot \prod_i \mathscr{M}_{\mathrm{hom},i}^{n_i}\;.
\end{equation}
Below, we provide the lowest order Hilbert bases for two discrete $R$ symmetries.

\subsection{\boldmath $R$ parity conserving  $\Z4^{R}$\unboldmath}
\label{sec:HilbertZ4R}
\noindent Inhomogeneous terms:
\begin{align*}
& (L \, \overline{E} \, \Hd)  ~,\quad  (Q \, \overline{D} \, \Hd)  ~,\quad  (Q \, 
\overline{U} \, \Hu)  ~,\quad  (L \, L \, \Hu \, \Hu)   ~,\quad   ( \overline{E}
\,  \overline{E} \, \Hd \, \Hd \, \Hd \, \Hd) ~,\\ &
  ( \overline{U} \, \overline{U} \, \overline{D} \, \overline{D} \, \overline{D}
\, \overline{D})   ~,\quad ( \overline{U} \, \overline{D} \, \overline{D} \, L
\, L \, \overline{E})   ~,\quad  (Q \, \overline{U} \, \overline{D} \,
\overline{D} \, \overline{D} \, L)  ~,\quad  (L \, L \, L \, L \, \overline{E}
\, \overline{E})   ~,\\ &
 (Q \, \overline{D} \, L \, L \, L \, \overline{E})
~,\quad  (Q \, Q \, \overline{D} \, \overline{D} \, L \, L)   ~,\quad   (Q \,
\overline{U} \, \overline{U} \, \overline{D} \, \overline{D} \, \overline{E} \,
\Hd)   ~,\quad (Q \, Q \, Q \,  Q  \, \overline{D} \, L\,\Hd)  ~,\\ & 
 (Q \, Q \,
Q \, L \, L \, \overline{E} \, \Hd) ~~,   ( \overline{U} \, \overline{U} \,
\overline{U} \, L \, \overline{E} \, \overline{E} \, \Hu)   ~,\quad (
\overline{D} \, \overline{D} \, \overline{D} \, L \, L \, L \, \Hu)  ~,\quad  (Q
\, Q \, Q \, Q \, \overline{U} \, L \, \Hu)   ~,\\ &
( \overline{U} \,
\overline{U} \, \overline{U} \, \overline{E} \, \overline{E} \, \overline{E} \,
\Hd \, \Hd) ~,\quad
(Q \, Q \, \overline{U} \, \overline{U} \, \overline{E} \,
\overline{E} \, \Hd \, \Hd)   ~,\quad   ( \overline{U} \, \overline{D} \,
\overline{D} \, \overline{D} \, \overline{D} \, \overline{D} \, \Hd \, \Hd) 
~,\\  &
(Q \, Q \, Q \, Q  \, \overline{U} \,  \overline{E} \, \Hd \, \Hd) 
~,\quad  (Q \, Q \, Q \, Q \, Q \, Q \, \Hd \, \Hd) ~,\quad    (Q \, \overline{U}
\,  \overline{U} \,  \overline{D} \, \overline{D} \, L \, \Hu \, \Hu)   ~,\\ &
(Q \, Q \, Q \, L \, L \, L \, \Hu \, \Hu)    ~,\quad ( \overline{U} \,
\overline{U} \, \overline{U} \, \overline{D} \, \overline{D} \, \overline{D} \,
\Hu \, \Hu)    
\;.
\end{align*}
\noindent Homogeneous terms:
\begin{multline*}
	 (\Hu \, \Hd)   ~,\quad  ( \overline{U} \, \overline{U} \, \overline{D} \, \overline{E})   ~,\quad 
(Q \, \overline{U} \, L \overline{E})   ~,\quad (Q \, Q \, \overline{U} \, \overline{D})   ~,\quad  (Q \, Q \, Q \, L)   ~,\quad ( \overline{D} \, \overline{D} \, \overline{D} \, L \, \Hd) ~~, \\
  ( \overline{U} \, \overline{D} \, \overline{D} \, L \, \Hu)   ~,\quad (L \, L \, L \, \overline{E} \, \Hu)   ~,\quad   (Q \, \overline{D} \, L \, L \, \Hu)  ~,\quad 
 (L \, L \, \overline{E} \, \overline{E} \, \Hd \, \Hd)   ~,\quad (Q \, \overline{D} \, L \, \overline{E} \, \Hd \, \Hd) ~~, \\
 (Q \, Q \, \overline{D} \, \overline{D} \, \Hd \, \Hd) ~,\quad    ( \overline{U} \, \overline{D} \, \overline{D} \, \overline{E} \, \Hd \, \Hd)   
~,\quad (Q  \, \overline{U} \, L \, \overline{E} \, \Hu \, \Hd)   ~,\quad (Q \, Q \, \overline{U} \, \overline{D} \, \Hu \, \Hd) ~~, \\
  (Q \, Q \, Q \, L \, \Hu \, \Hd)  ~,\quad  (Q \, Q \, \overline{U} \,  \overline{U} \, \Hu \, \Hu)  ~,\quad  ( \overline{U} \, \overline{U} \,  \overline{U} \, \overline{E} \, \Hu \, \Hu)  
 ~,\quad  (Q \, \overline{U} \,  \overline{E} \, \overline{E} \, \Hd \, \Hd \, \Hd) ~~, \\
 (Q \, Q \, Q \,  \overline{E} \, \Hd \, \Hd \, \Hd)   ~,\quad (L \, L \, L \, \overline{E} \, \Hu \, \Hu \, \Hd)  ~,\quad   (Q \, \overline{D} \, L \, L \, \Hu \, \Hu \, \Hd)  ~~, \\
 (Q \, \overline{U} \, L \, L \, \Hu \, \Hu \, \Hu) ~,\quad 
 (Q \, Q \, Q \, L \, \Hu \, \Hu \, \Hd \, \Hd)  ~,\quad   (L \, L \, L \, L \, \Hu \, \Hu \, \Hu \, \Hu)  \;. \\
\end{multline*}

\subsection{\boldmath Non--perturbative RPV  $\Z3^{R}$\unboldmath}
\label{sec:HilbertZ3R}

\noindent Inhomogeneous terms:
\begin{align*}
& (L \, \overline{E} \, \Hd)  ~,\quad  (Q \, \overline{U} \, \Hu)  ~,\quad  (Q \, \overline{D} \, \Hd) ~,\quad   (L \, L \, \Hu \, \Hu)  
~,\quad   (Q \, Q \, Q \, Q \overline{U}) ~,\quad   (Q \, Q \, Q \, L \, L \, \Hu) ~,\\ &
 (Q  \, Q  \, \overline{U}  \, \overline{U}  \, \overline{E})  ~,\quad    (Q  \, Q  \, Q  \, L   \, \overline{E}  \, \Hd  \, \Hd)  
~,\quad  (Q  \, \overline{U}  \, L  \, L   \, \overline{E}  \, \Hu)  
~,\quad  (Q  \, Q  \,  \overline{U}  \,  \overline{D}  \, L  \, \Hu) ~~ , ~~  ( \overline{U}  \, \overline{U}  \, \overline{U}  \, \overline{E}  \, \overline{E}) ~,\\ &
 ( \overline{D}  \, \overline{D}  \, \overline{D}  \, L  \, L) ~,\quad    (Q  \, Q  \, Q  \, Q  \, Q  \, Q  \, L  \, L)   ~,\quad (Q  \, Q  \, Q  \, L  \, L  \, \Hu  \, \Hu  \, \Hd) 
~,\quad     ( \overline{U}  \, \overline{U}  \,  \overline{D}  \, L  \, \overline{E}  \, \Hu) ~,\\ &
 (Q  \, \overline{U} \, \overline{U}  \, \overline{D}   \, \overline{D}  \, \Hu)  ~,\quad  (Q  \, Q  \, Q  \, Q  \, \overline{U}  \, L \, L \, \overline{E})  ~,\quad    
(Q  \, Q \, Q \, Q \, Q \, \overline{U} \, \overline{D} \, L) ~,\quad
( \overline{D} \, \overline{D} \, \overline{D} \, L \, L \, \Hu \, \Hd) ~,\\ &
 (Q  \, Q \, \overline{U}  \, \overline{U}  \, L  \, L \overline{E} \, \overline{E}) ~,\quad   (Q \, Q \, Q \, Q \, \overline{U} \, \overline{U}  \, \overline{D} \, \overline{D}) ~,\quad 
    (Q \, Q \, Q \, \overline{U} \, \overline{U} \, \overline{D} \, L \overline{E}) ~,\quad  (Q \, Q \, Q \, L \, L \, L \, L \, \overline{E} \, \Hu)\;.
\end{align*}
\noindent Homogeneous terms:
\begin{multline*}
	(\Hu \, \Hd)   ~,\quad  (Q \, Q \, Q \, \Hd)  ~,\quad   (L \, L  \, \overline{E})  ~,\quad   (Q \, \overline{D} \, L)  ~,\quad   (Q \,  \overline{U} \,  \overline{E} \, \Hd)   
~,\quad    ( \overline{U}  \, \overline{D} \,  \overline{D}) ~~,  \\
 (L \, L  \, \overline{E} \, \Hu \, \Hd) ~,\quad  (Q \, Q \, Q \, \Hu \, \Hd \, \Hd) ~,\quad   (Q  \, \overline{U} \, L \, \Hu \, \Hu)  
 ~,\quad    (Q  \, \overline{D} \, L \, \Hu \, \Hd) ~~, \\
 (L \, L \, L \, \Hu \, \Hu \, \Hu)   ~,\quad ( \overline{D} \,  \overline{D}  \, \overline{D} \, \Hd \, \Hd) ~,\quad 
( \overline{U} \,  \overline{U} \,  \overline{D} \, \Hu \, \Hu) ~,\quad   (Q \, Q \, Q \, L \, L \overline{E} \, \Hd)  ~~,  \\
 (Q \, Q \, Q \, Q \,  \overline{U} \, L \, \Hu)  ~,\quad   (Q \, Q \, Q \, Q \,  \overline{D} \, L \, \Hd)  ~,\quad 
(Q \, Q \, Q \, L \, L \, L \, \Hu \, \Hu) ~,\quad      (Q \,  \overline{U} \, L \, L \,  \overline{E} \,  \overline{E} \, \Hd) ~~,  \\ 
  (Q \, Q \,  \overline{U} \,  \overline{U} \, L  \, \overline{E} \, \Hu)  ~,\quad   (Q \, Q \, Q  \, \overline{U}  \, \overline{U}  \, \overline{D} \, \Hu)    ~,\quad 
(Q \, Q \, Q \,  \overline{U} \,  \overline{D} \,  \overline{D} \, \Hd)   ~,\quad  (Q \, Q  \, \overline{U} \,  \overline{D} \, L \,  \overline{E} \, \Hd)\;.
\end{multline*}

\bibliography{Orbifold}
\addcontentsline{toc}{section}{Bibliography}
\bibliographystyle{NewArXiv} 
\end{document}